\documentclass[twoside]{elsart}
\usepackage{epsfig}
\usepackage{array}
\usepackage{amssymb,amsbsy,amsmath}

\def\bra#1{\langle \, {#1} \, | \;}
\def\ket#1{\; | \, {#1} \, \rangle}
\newcommand{\braket}[2]{\langle \, {#1} \, | \, {#2} \, \rangle}
\def\erw#1{\,\langle \, {#1} \, \rangle\,}
\newcommand{\vek}[1]{{\!\vec{\,#1}}}

\newcommand{\fm}{\mbox{$\,$fm}}
\newcommand{\MeV}{\mbox{$\,$MeV}}
\newcommand{\element}[2]{$^{#1}$#2}
\def\leap{\raisebox{-.6ex}{$\stackrel {<}{\sim}$}} 
\newcommand{\op}[1]{#1}
\newcommand{\coop}[1]{\widehat{#1}}

\newcommand{\pp}[2]{\frac{\partial \, {#1}}{\partial \, {#2}}\;}

\newcommand{\dint}{\mbox{d}}
\renewcommand{\half}{\frac{1}{2}}

\newcommand{\RPM}[1]{R_{\pm}(#1)}
\newcommand{\RM}[1]{R_{-}(#1)}
\newcommand{\RMPr}[1]{R_{-}^{\,\prime}(#1)}

\newcommand{\RP}[1]{R_{+}(#1)}
\newcommand{\RPPr}[1]{R_{+}^{\,\prime}(#1)}

\newcommand{\DRM}[1]{R_{-}^{\,\prime}(#1)}
\newcommand{\DRP}[1]{R_{+}^{\,\prime}(#1)}
\newcommand{\DDRP}[1]{R_{+}^{\,\prime\prime}(#1)}
\newcommand{\DDDRP}[1]{R_{+}^{\,\prime\prime\prime}(#1)}
\newcommand{\vecrdr}{\frac{\vek{r}}{r}}
\newcommand{\xref}[1]{\protect\ref{#1}}
\newcommand{\fmref}[1]{(\protect\ref{#1})}
\newcommand{\figref}[1]{fig.~\protect\ref{#1}}
\newlength{\CaptionWidth}
\newcommand{\mycaption}[2]{%
\begin{center}\begin{minipage}{\CaptionWidth}%
\caption[]{#1}\label{#2}%
\end{minipage}\end{center}%
}%

\begin{document}
%
\typeout{   --- >>>   correlator paper   <<<   ---   }
\typeout{   --- >>>   correlator paper   <<<   ---   }
\typeout{   --- >>>   correlator paper   <<<   ---   }
\begin{frontmatter}
\title{A unitary correlation operator method}
 
\author{H. Feldmeier\thanksref{HF}}
\author{, T. Neff\thanksref{TN}}
\author{, R. Roth\thanksref{RR}}
\author{ and J. Schnack\thanksref{JS}}
\address{Gesellschaft f\"ur Schwerionenforschung mbH, \\ 
         Postfach 110 552, D-64220 Darmstadt \&\\
         Technische Hochschule Darmstadt}

\thanks[HF]{email: h.feldmeier\char'100gsi.de,
            WWW:~http://www.gsi.de/$\sim$feldm}
\thanks[TN]{email: t.neff\char'100gsi.de,
            WWW:~http://www.gsi.de/$\sim$tneff}
\thanks[RR]{email: r.roth\char'100gsi.de,
            WWW:~http://www.gsi.de/$\sim$rroth}
\thanks[JS]{email: j.schnack\char'100gsi.de,
            WWW:~http://www.gsi.de/$\sim$schnack}

\begin{abstract}

\noindent
The short range repulsion between nucleons is treated by a
unitary correlation operator which shifts the nucleons away from
each other whenever their uncorrelated positions are within the
replusive core.  By formulating the correlation as a
transformation of the relative distance between particle pairs,
general analytic expressions for the correlated wave functions and
correlated operators are given.  The decomposition of correlated
operators into irreducible $n$--body operators is discussed.
The one-- and two--body--irreducible parts are worked out
explicitly and the contribution of three--body correlations is
estimated to check convergence.  Ground state energies of nuclei
up to mass number $A=48$ are calculated with a
spin--isospin--dependent potential and single Slater
determinants as uncorrelated states.  They show that the deduced
energy-- and mass--number--independent correlated two--body
Hamiltonian reproduces all "exact" many--body calculations
surprisingly well.

\vspace{1ex}

\noindent{\it PACS:} 
21.60.-n;        
21.30.Fe;        
21.45.+v;        
13.75.C          

\vspace{1ex}

\noindent{\it Keywords:} Few--body systems; Fermion systems;
Ground state properties; Short range correlations;
Nucleon-nucleon interactions; Effective interactions;
Cluster approximation
\end{abstract}
\end{frontmatter}
\newpage
\tableofcontents
\newpage
\raggedbottom
\section{Introduction}

In many--body physics the most convenient trial states are
products of single--particle states which for identical bosons
or fermions are symmetrized or antisymmetrized with respect to
particle exchange. An adequately selected set of those product
states spans a subspace of the Fock space called model space. In
the following we shall use the term product state for bosons and
fermions as well and proper statistics is implied without
mentioning.

For particles with a rigid internal structure like atoms or
nucleons the two--body interaction between the particles
contains a short--ranged repulsive core. For this situation
product states are not appropriate because a numerically
tractable number of them can usually not represent the strong
depletion of probability in the two--body density distribution
at short distances between the particles. Or in other words, the
true eigenstates of the exact Hamiltonian contain components
which lie outside the model space in the so--called $Q$--space.

There are in general two types of methods to remedy the incompatibility
of the Hamiltonian with the trial states which span the model
space \cite{AHP93}. The first is to adapt the Hamiltonian
by replacing the interaction with an effective one which takes 
the scattering between model space and $Q$--space into account. 
The Brueckner $G$--Matrix formalism is of this type
\cite{Bru55}. The method of effective operators allows 
to calculate the whole spectrum of low lying states 
including transitions between them.

The second way is to introduce the short--range repulsive
correlations into the many--body state by applying correlation
operators \cite{BPP93} onto the trial product state. In the
simplest case it is a multiplication with Jastrow correlation
functions \cite{Jas55}. This approach is mainly used to find
ground--state energies by minimizing with respect to the
parameters contained in the correlated trial state. Spectra and
transitions cannot be calculated easily in this scheme.

In this article we propose the Unitary Correlation Operator
Method (UCOM), which to a certain extend combines the advantages
of the above mentioned schemes.  It correlates the product
state, $\ket{\Phi}$, by means of a unitary correlator,
$\op{C}$. The identity
\begin{equation}
\bra{\Psi}\op{H}\ket{\Psi^\prime} 
= 
\bra{\Phi}\op{C}^{\dagger}\,\op{H}\,\op{C}\ket{\Phi^\prime}
= 
\bra{\Phi}\op{C}^{-1}\,\op{H}\,\op{C}\ket{\Phi^\prime}
=
\bra{\Phi}\coop{H}\ket{\Phi^\prime}
\end{equation}
shows that one may either speak of the original ``untamed''
Hamiltonian $\op{H}$ sandwiched between correlated states
$\ket{\Psi}=\op{C}\ket{\Phi}$ or of a correlated Hamiltonian
$\op{C}^{-1}\,\op{H}\,\op{C}$ represented in the model space of
the uncorrelated product states $\ket{\Phi}$. The task of the
correlator $\op{C}$ is to ``tame'' the short--ranged repulsive
part of $\op{H}$ by means of a unitary transformation. This
general idea exists for a long time \cite{PrS64,BaK73,SOK94} but
has not been pursued very much.

When the correlator $\op{C}$ is applied to the Hamiltonian the method can be
regarded as a pre-diagonalization such that large matrix
elements connecting model space and $Q$--space are removed. 
$\coop{H}=\op{C}^{-1}\,\op{H}\,\op{C}$ can then for example be
diagonalized in a shell model configuration space in order to
describe long--range many--body correlations in low lying
eigenstates. In this sense it is similar to the method of
effective operators, although there a projection onto the model
space is used rather than a unitary transformation.

When $\op{C}$ acts on the uncorrelated state $\ket{\Phi}$ one
has more the picture of a trial ansatz for the many--body state
as for example used in Jastrow--type methods. However, the
advantage of a unitary $\op{C}$ is that the norms
$\braket{\Psi}{\Psi}=\bra{\Phi}\op{C}^\dagger\,\op{C}\ket{\Phi}=\braket{\Phi}{\Phi}$
of correlated and uncorrelated states are identical so that
there is no need to develop expansion methods when 
calculating the denominator in expectation values like
${\bra{\Phi}\op{C}^{\dagger}\op{H}\op{C}\ket{\Phi}}
/{\bra{\Phi}\op{C}^{\dagger}\op{C}\ket{\Phi}}$ \cite{Cla79}.

We propose to write $\op{C}$ as a generalized shift operator,
which moves two particles away from each other whenever they are
within the classically forbidden region of the repulsive core.
In section \xref{Sec-2-0} we discuss the many--body nature of
the correlator $\op{C}$ and of the correlated operators and give
analytic expressions for $\op{C}$ and the correlated Hamiltonian
in terms of a coordinate transformation of the relative
distance. In the two--body space the resulting correlated
Hamiltonian has no repulsive core anymore, instead a
non-locality occurs at short distances which is written in terms
of a quadratic momentum dependence with a reduced mass which is
a function of the distance. The eigenstates, bound as well as
scattering states, of the correlated Hamiltonian coincide with
the exact eigenstates of the original Hamiltonian outside the
range of the correlator, therefore the two interactions are
completely phase--shift equivalent \cite{Eks60,CCD70}.

For the many--body system our aim is to devise a correlator
which does not depend on energy and particle number so that the
same correlated Hamiltonian $\coop{H}$ can be used for many
nuclei. In section \xref{Sec-4-0} results will be given for mass
numbers ranging from $A=2$ to $A=48$. The main
problem consists in separating the variational degrees of
freedom residing in the correlator $\op{C}$ from those in the
product state $\ket{\Phi}$ such that the irreducible three-- and
higher--body parts of $\coop{H}$ remain small.  These aspects
are discussed in section \xref{Sec-2-4} and in section
\xref{Sec-3-0}, where the concept is applied to
spin--isospin--independent forces and small nuclei first.  The
convergence of the expansion of the correlated Hamiltonian into
one--, two-- and three--body parts is discussed in terms of a
smallness parameter which is the product of the 
density times a correlation volume.

Then in section \xref{Sec-4-0} the correlator is extended to
spin--isospin--dependent forces which are more realistic for
heavier nuclei. We calculate ground state energies up to
\element{48}{Ca} and compare them to existing ``exact''
results. The agreement is astonishing considering that the
uncorrelated many--body states are single Slater determinants.

\section{The concept of a unitary correlator}
\label{Sec-2-0}

In this section the advantage of representing improved trial
states by a unitary transformation of product states will be
elucidated in a quite general way. The following subsections
will explain in detail the explicit form of the correlator
$\op{C}$ and the expansion of a correlated operator into
$n$--body operators, in particular the expansion of the correlated
Hamiltonian.

Throughout this section we use the convention that $\ket{\Psi}$
or $\ket{\psi}$ stand for correlated states and $\ket{\Phi}$ or
$\ket{\phi}$ for uncorrelated states.  The states $\ket{\Phi_n}$
denote a basis set of many--body product states (symmetrized or
antisymmetrized if necessary).

The eigenstates $\ket{n}$ of the many--body eigenvalue problem
\begin{eqnarray}
\op{H} \ket{n}
=
E_n \ket{n}
\end{eqnarray}
can be written as
\begin{eqnarray}
\ket{n}
=
\op{C}[\op{H},\Phi] \; \ket{\Phi_n}
\end{eqnarray}
where $\op{C}\left[\op{H},\Phi\right]$ is the unitary operator
\begin{eqnarray}\label{E-2-0-A}
\op{C}[\op{H},\Phi]
=
\sum_k \;
\ket{k}\bra{\Phi_k}
\end{eqnarray}
which diagonalizes the Hamiltonian $\op{H}$.

The idea is to devise a unitary correlation operator $\op{C}$
which approximates $\op{C}[\op{H},\Phi]$ for low
energies, i.e. for states $\ket{k}$ with $E_k$ in the vicinity
of the ground state. From \fmref{E-2-0-A} it is clear that
$\op{C}$ depends on the choice of the product states
$\ket{\Phi_k}$ and on the Hamiltonian $\op{H}$ via its
eigenstates $\ket{k}$.

Since there is a certain amount of freedom in choosing the
single--particle states which build the product states
$\ket{\Phi_n}$ there is no need to be so ambitious to really
approximate $\op{C}[\op{H},\Phi]$ in the sense of
an operator equality
$\op{C}=\op{C}[\op{H},\Phi]$. One rather divides the
burden among $\op{C}$ and $\ket{\Phi_n}$ in such a way that
$\ket{\Phi_n}$ or linear combinations of a small number of
$\ket{\Phi_n}$ describe the long--range correlations (mean field,
deformations) while $\op{C}$ takes care of the short--range
repulsive correlations.

In other words the physical task of $\op{C}$ is to provide a
mapping between the Hilbert space spanned by low lying
correlated states $\ket{k}$ and the Hilbert space of 
uncorrelated product states $\ket{\Phi_k}$ which do not 
contain (repulsive) short--range correlations.

One main advantage of a unitary correlator $\op{C}$ is that the
correlated state
\begin{eqnarray}
\ket{\Psi} = \op{C} \ket{\Phi}
\end{eqnarray}
has the same norm as the uncorrelated state $\ket{\Phi}$
so that the norm in the denominator of an expectation value
\begin{eqnarray}\label{E-2-0-B}
\bra{\Psi}\op{B}\ket{\Psi} 
= 
\frac{\bra{\Phi}\op{C}^\dagger\, \op{B}\; \op{C}\ket{\Phi}}
     {\bra{\Phi}\op{C}^\dagger \op{C}\ket{\Phi}}
\end{eqnarray}
does not require special treatment as for example in the
Jastrow method \cite{Jas55,Cla79}.

If the correlator $\op{C}$ is chosen to be state independent one
may equivalently use correlated states $\ket{\Psi}$ and uncorrelated
operators $\op{B}$ or uncorrelated states $\ket{\Phi}$ and
correlated operators $\coop{B}$
\begin{eqnarray}
\bra{\Psi}\op{B}\ket{\Psi^\prime} = \bra{\Phi}\coop{B}\ket{\Phi^\prime}
\ ,
\end{eqnarray}
where the correlated operator
\begin{eqnarray}
\coop{B} = \op{C}^\dagger\, \op{B}\; \op{C} 
= \op{C}^{-1}\, \op{B}\; \op{C}
\end{eqnarray}
does not dependent on the particular states
$\ket{\Phi}$ and $\ket{\Phi^\prime}$.

If $\op{C}$ is not unitary as for example in the Jastrow method
\cite{Jas55} one may from eq. \fmref{E-2-0-B} also define a
correlated operator $\coop{B}_{\mbox{\scriptsize Jastrow}}$ as
\begin{eqnarray}
\coop{B}_{\text{Jastrow}}
=
\frac{\op{C}^\dagger_{\text{Jastrow}}\, \op{B}\; \op{C}_{\text{Jastrow}}}
     {\bra{\Phi}\op{C}^\dagger_{\text{Jastrow}}
     \op{C}_{\text{Jastrow}} \ket{\Phi}}
\ ,
\end{eqnarray}
but this depends on the state $\ket{\Phi}$.

For a unitary correlator $\op{C}$ there is the duality that one
may regard $\op{C}$ as a part of the trial state
$\op{C}\ket{\Phi}$ which provides further variational degrees of
freedom beyond those already contained in $\ket{\Phi}$. Or one
may view $\op{C}$ as a unitary transformation of the Hamiltonian
which amounts to a pre-diagonalization
\begin{eqnarray}
\bra{\Psi_k}\op{H}\ket{\Psi_l} 
=
\bra{\Phi_k}\op{C}^{-1}\, \op{H}\; \op{C}\ket{\Phi_l}
= 
\bra{\Phi_k}\coop{H}\ket{\Phi_l}
\ .
\end{eqnarray}
This means that the transformed Hamiltonian
$\coop{H}=\op{C}^{-1}\,\op{H}\;\op{C}$ is much better
represented in the product basis $\ket{\Phi_n}$ than the
original Hamiltonian.

Both ways of viewing the correlator $\op{C}$ are mathematically
equivalent. The first, where $\op{C}$ is part of the trial state,
is more suited for the Ritz variational principle where one
wants to find only the ground state and its energy. The second
is better adopted to the language of multi--configuration mixing
and effective operators. In any case the physics contained in
$\op{C}$ is the same, namely to describe the short range
correlations.

\subsection{Decomposition of correlated operators into $n$--body operators}
\label{Sec-2-1}

In order to ensure unitarity the correlator $\op{C}$ is written as
\begin{eqnarray}
\op{C} = \exp\{-i \op{G}\} \ ,\qquad
\op{G} = \op{G}^\dagger
\ ,
\end{eqnarray}
where $\op{G}$ is the hermitean generator of the
correlations. Because we treat fermion systems in this work,
particle number conserving generators are considered only. 
Furthermore the generator $\op{G}$ is invariant
under permutations of the particles. $\op{G}$ has to be a
two--body operator or higher because a one--body operator would only
cause a unitary transformation of the single--particle
states and this variational degree of freedom is already
present in the product--state $\ket{\Phi}$.
\begin{eqnarray}
\op{G} = \sum_{i<j}^A\;\op{g}(i,j) 
+\mbox{three--body}
+\mbox{higher}
\end{eqnarray}
Any correlated operator
\begin{eqnarray}
\coop{B} = \op{C}^{-1}\; \op{B}\; \op{C}
=
\exp\{+i \op{G}\}\;
\op{B}\;
\exp\{-i \op{G}\}
\end{eqnarray}
will therefore be in general a superposition of zero--, one--,
two--, three-- and higher--particle operators.

In order to avoid confusion we will use the following
notation. An irreducible $n$--body operator $\op{B}$ in an 
$A$--particle space is denoted as
\begin{eqnarray}
\op{B}^{[n]}_A
=
\sum_{i_1 < \cdots < i_n}^A \op{b}^{[n]}(i_1,\cdots,i_n)
\ ,
\end{eqnarray}
where $\op{b}^{[n]}(i_1,\cdots,i_n)$ is the actual operator with
the $n$ particle indices $i_1,\cdots, i_n$. If $n>A$ then
$\op{B}^{[n]}_A=0$. For example, the Hamiltonian which consists
of a one--body kinetic energy $\op{T}$ and a two--body potential
$\op{V}$ is written in this elaborate way as
\begin{eqnarray}
\op{H} = \op{T} + \op{V}
\equiv
\op{H}_A = \op{T}^{[1]}_A + \op{V}^{[2]}_A
=
\sum_{i}^A \op{t}^{[1]}(i)
+
\sum_{i<j}^A \op{v}^{[2]}(i,j)
\ .
\end{eqnarray}
A missing superscript $[n]$ means that the operator is a
combination of different many--body operators.
Of course later on when there is no danger of confusion, superscripts and
subscripts are omitted again.

The decomposition of the correlated operator
$\coop{B}\equiv\coop{B}_A $ into a sum of irreducible $n$--body
operators is written as
\begin{eqnarray}\label{E-2-1-A}
\coop{B}_A 
\equiv 
\op{C}_A^{-1}\, \op{B}_A\; \op{C}_A
=
\sum_{n=0}^A \coop{B}^{[n]}_A
\ ,
\end{eqnarray}
where in the $A$--body space the $n$--body operator
\begin{eqnarray}\label{E-2-1-E}
\coop{B}^{[n]}_A
=
\sum_{i_1 < \cdots < i_n}^A \coop{b}^{[n]}(i_1,\cdots,i_n)
\qquad\mbox{for}\quad n \geq 1
\end{eqnarray}
is given by
\begin{eqnarray}\label{E-2-1-B}
\coop{b}^{[n]}
&:=&
\coop{B}_n
-
\sum_{k=1}^{n-1} \coop{B}^{[k]}_n
\\
&=&
\op{C}^{-1}_n\, \op{B}_n\; \op{C}_n
-
\sum_{k=1}^{n-1} 
\sum_{i_1 < \cdots < i_k}^n 
\coop{b}^{[k]}(i_1,\cdots,i_k)
\nonumber
\ .
\end{eqnarray}
The first non-vanishing $\coop{b}^{[n]}$ depends on how many
particles the operator $\op{B}$ connects. The trivial case is 
when $\op{B}$ is proportional to the unit operator then
$\coop{B}_A=\op{C}_A^{-1}\op{B}_A\op{C}_A=\coop{B}^{[0]}_A=\op{B}_A$.
If $\op{B}=\sum_{i_1<\cdots<i_m}^A\op{b}^{[m]}(i_1,\cdots,i_m)$
is an $m$--body operator then the first non-vanishing
contribution comes from $\coop{b}^{[m]}$ and is given by
\begin{eqnarray}
\coop{b}^{[m]}
=
\op{C}^{-1}_m\, \op{b}^{[m]}\; \op{C}_m
\ .
\end{eqnarray}
The decomposition \fmref{E-2-1-A} is often called ``cluster
expansion''. This is at this stage somewhat misleading as the
particles $i_1,i_2,\cdots,i_n$ whose coordinates show up in
$\coop{b}^{[n]}(i_1,\cdots,i_n)$ need not be close to each
other. 

From now on we assume the generator to be a two--body operator
$\op{G}=\sum_{i<j}^A\op{g}(i,j)$.
Of particular interest is the effect of the correlation on one--
and two--body operators. For example the kinetic energy, a
one--body operator, transforms to
\begin{eqnarray}
\coop{T}_A 
\equiv 
\op{C}^\dagger_A\; \op{T}_A\; \op{C}_A
=
\coop{T}^{[1]}_A + \coop{T}^{[2]}_A + \coop{T}^{[3]}_A + \cdots
\ ,
\end{eqnarray}
where the one--body part is just the regular uncorrelated
kinetic energy
\begin{eqnarray}
\coop{T}^{[1]}_A
=
\sum_{i}^A \coop{t}^{[1]}(i)
=
\sum_{i}^A \op{t}^{[1]}(i)
=
\op{T}^{[1]}_A
\end{eqnarray}
because the generator is a two--body operator. The two--body part
$\coop{T}^{[2]}_A$ is obtained from the general
eqs. \fmref{E-2-1-E} and \fmref{E-2-1-B} as
\begin{eqnarray}
&& \coop{T}^{[2]}_A
=
\sum_{i<j}^A \coop{t}^{[2]}(i,j)\qquad\text{with}
\\[2mm]
&& \coop{t}^{[2]}(i,j)
=
\op{c}(i,j)^{-1} 
\big(\op{t}^{[1]}(i) + \op{t}^{[1]}(j)\big)\;
\op{c}(i,j)
-
\big(\op{t}^{[1]}(i) + \op{t}^{[1]}(j)\big)
\nonumber
\end{eqnarray}
and $\op{c}(i,j) = \exp\{-i \op{g}^{[2]}(i,j) \}$ denotes the
correlator between particle $i$ and $j$.
The obvious meaning is that for each pair $(i,j)$ the two--body
part is the difference between correlated and uncorrelated 
kinetic energy for the pair. The three--body part of the
correlated kinetic energy operator is according to
eqs. \fmref{E-2-1-E} and \fmref{E-2-1-B} given by
\begin{eqnarray}\label{E-2-1-C}
&&
\coop{T}^{[3]}_A
=
\sum_{i<j<k}^A \coop{t}^{[3]}(i,j,k)\qquad\text{and} 
\\[2mm]
&&
\coop{t}^{[3]}(i,j,k)
=
\op{c}(i,j,k)^{-1} \;
\big(\op{t}^{[1]}(i) + \op{t}^{[1]}(j) + \op{t}^{[1]}(k)\big) \;
\op{c}(i,j,k)
\nonumber
\\
&&\qquad\qquad\quad
-
\big(\coop{t}^{[2]}(i,j) + \coop{t}^{[2]}(i,k) +
\coop{t}^{[2]}(j,k)\big)
\nonumber
\\
&&\qquad\qquad\quad
-
\big(\op{t}^{[1]}(i) + \op{t}^{[1]}(j) + \op{t}^{[1]}(k)\big)
\nonumber
\end{eqnarray}
with $\op{c}(i,j,k) = \exp \{
-i(\op{g}^{[2]}(i,j)+\op{g}^{[2]}(i,k)+\op{g}^{[2]}(j,k)) \}$.
The three--body part correlates all triples $(i,j,k)$ of the
system. But it is important to note that only the genuine
irreducible three--body correlations appear in \fmref{E-2-1-C}
because all two--body correlations and the uncorrelated kinetic
energy are subtracted out.

When transforming a two--body operator
$\op{V}^{[2]}_A=\sum_{i<j}\op{v}^{[2]}(i,j)$ like the potential,
the correlated operator
\begin{eqnarray}
\coop{V}_A 
\equiv 
\op{C}^{-1}_A\; \op{V}_A\; \op{C}_A
=
\coop{V}^{[2]}_A + \coop{V}^{[3]}_A + \cdots
\end{eqnarray}
starts with a two--body part which is already correlated
\begin{eqnarray}
&&\coop{V}^{[2]}_A
=
\sum_{i<j}^A \coop{v}^{[2]}(i,j)\qquad\mbox{with}
\\[2mm]
&&\coop{v}^{[2]}(i,j)
=
\op{c}(i,j)^{-1} \;
\op{v}^{[2]}(i,j) \;
\op{c}(i,j) 
\nonumber
\ .
\end{eqnarray}
This means that the bare interaction $\op{v}^{[2]}$ is in
lowest order replaced by the correlated interaction
$\coop{v}^{[2]}$ which will be much less repulsive at short
distances.

The specific form of eq. \fmref{E-2-1-B} for the three--body
part is
\begin{eqnarray}\label{E-2-1-D}
&&
\coop{V}^{[3]}_A
=
\sum_{i<j<k}^A \coop{v}^{[3]}(i,j,k)\qquad\mbox{with} 
\\[2mm]
&&
\coop{v}^{[3]}(i,j,k)
=
\op{c}(i,j,k)^{-1} \;
\big(\op{v}^{[2]}(i,j) + \op{v}^{[2]}(i,k) +
\op{v}^{[2]}(j,k)\big) \;
\op{c}(i,j,k)
\nonumber
\\
&&\qquad\qquad\quad
-
\big(\coop{v}^{[2]}(i,j) + \coop{v}^{[2]}(i,k) +
\coop{v}^{[2]}(j,k)\big)
\nonumber
\ .
\end{eqnarray}
From the explicit way of writing eqs. \fmref{E-2-1-C} and
\fmref{E-2-1-D} it is obvious that
the three--body part will only contribute appreciably if the
probability of finding three particles simultaneously in the
correlation volume is high. 
The importance of the three--body correlations will be discussed
in more detail in section \xref{Sec-3-3}.

Another very important issue is the ``cluster decomposition
property''. If a system decomposes into locally separated
subsystems with particle numbers $A, B,\dots,Z$, which means that
the uncorrelated state can be written as
\begin{eqnarray}
\ket{\Phi}
=
\ket{\Phi_A}\otimes\ket{\Phi_B}
\otimes\cdots\otimes
\ket{\Phi_Z}
\end{eqnarray}
where $\ket{\Phi_A}$, $\ket{\Phi_B}$ and so on are locally
disconnected, then
\begin{eqnarray}
\op{C}\ket{\Phi}
=
\op{C}_A\ket{\Phi_A}\otimes\op{C}_B\ket{\Phi_B}
\otimes\cdots\otimes
\op{C}_Z\ket{\Phi_Z}
\ .
\end{eqnarray}
From the construction of $\op{C}$, which is
\begin{eqnarray}
\op{C}
=
\exp\!\Big\{ -i \sum_{i<j}^{A+B+\cdots+Z}\op{g}^{[2]}(i,j) \Big\}
\ ,
\end{eqnarray}
the validity of the decomposition property is obvious because
first, $\op{g}^{[2]}(i,j)=0$ if $i$ and $j$ do not belong to the same
subsystem and second, $[\op{g}^{[2]}(i,j),\op{g}^{[2]}(k,l)]=0$ if 
$(i,j)\in A$ and $(k,l)\not\in A$. Therefore, $\op{C}$
can be decomposed as
\begin{eqnarray}
\op{C}\ket{\Phi}
&=&
\exp\!\Big\{ -i \sum_{i<j}^{A}\op{g}^{[2]}(i,j) \Big\}
\ket{\Phi_A}
\otimes\cdots
\\
&&\qquad\qquad\qquad\qquad\qquad
\cdots\otimes
\exp\!\Big\{ -i \sum_{i<j}^{Z}\op{g}^{[2]}(i,j) \Big\}
\ket{\Phi_Z}
\nonumber
\end{eqnarray}
and hence the ``cluster decomposition property'' is fulfilled.

It is easy to see that the cluster decomposition property of the
correlator $\op{C}$ is carried over to each irreducible
$n$--body  term in $\coop{B}$ (see eq. \fmref{E-2-1-A}).

\subsection{Representation as coordinate transformation}

The repulsion at short distances will keep the particles from
getting too close to each other. This feature is not included in
a (symmetrized/antisymmetrized) product state where the
probability to find two particles (with different spin) at the
same place is just the product of the one--body
densities ($\pm$ exchange term). Therefore, the task of the 
correlator is to shift two particles away from each other 
whenever they are close.

In this section we define a generator $\op{g}(1,2)$ which is
producing an outward radial shift depending on the uncorrelated
distance of the particles. We express the action of the
correlator in the two--body space
$\op{c}(1,2)=\exp\{-i\op{g}(1,2)\}$ in terms of a
coordinate transformation $R_{\pm}(r)$ of the radial distance
$r$.

The relative distance and the relative momentum for equal mass
particles are denoted by
\begin{eqnarray}
\vek{r} = \vek{x}(1) - \vek{x}(2)
\ ,\qquad
\vek{q} = \frac{1}{2}\left(\vek{p}(1) - \vek{p}(2)\right)
\ ,
\end{eqnarray}
respectively.  A generator $\op{g}(1,2)=\op{g}(\vek{r},\vek{q})$
which creates a position--dependent shift may be written in the
hermitean form
\begin{eqnarray}\label{E-2-2-A}
\op{g}(\vek{r},\vek{q}) 
= 
\frac{1}{2}\left\{
\left(\vek{q}\cdot\frac{\vek{r}}{r}\right)\;s(r)
+ s(r)\;\left(\frac{\vek{r}}{r}\cdot\vek{q}\right)\right\}
\ ,\qquad
r\equiv|\vek{r}|
\ .
\end{eqnarray}
One expects the unitary operator
$\op{c}(1,2)=\exp\{-i\op{g}(\vek{r},\vek{p})\}$ to shift a
relative position $\vek{r}$ to about
$\vek{r}+s(r)\frac{\vek{r}}{r}$. The exact transformation will
be derived below.  The idea is to find a suited function $s(r)$
such that $\op{g}(\vek{r},\vek{p})$ moves the probability
amplitude out of the classically forbidden region of the
repulsion.

In coordinate representation the action of the generator
$\op{g}(\vek{r},\vek{q})$ on a relative wave function
$\braket{\vek{r}}{\phi}$ is given by
\begin{eqnarray}
\bra{\vek{r}}\op{g}\ket{\phi}
&=& -i\left(\half\pp{s}{r} + \frac{s}{r} + s(r)\pp{}{r}\right)
\braket{\vek{r}}{\phi}
\\
&=&
-i\frac{1}{r\sqrt{s(r)}}\;\;s(r)\pp{}{r}\;\;r\sqrt{s(r)}\;
\braket{\vek{r}}{\phi}
\ .
\end{eqnarray}
From that we obtain the coordinate representation of the correlator
\begin{eqnarray}
\bra{\vek{r}}\op{c}\ket{\phi}
&=&
\exp\left\{
-\frac{1}{r\sqrt{s(r)}}\;s(r)\pp{}{r}\;r\sqrt{s(r)}
\right\}\;
\braket{\vek{r}}{\phi}
\\
\label{E-2-2-C}
&=&
\frac{1}{r\sqrt{s(r)}}\;
\exp\left\{-s(r)\pp{}{r}\right\}
\;r\sqrt{s(r)}\;
\braket{\vek{r}}{\phi}
\ .
\end{eqnarray}
The last line can be verified using the power expansion of the
exponential. Writing $\op{c}$ in this way suggests a coordinate 
transformation
\begin{eqnarray}\label{E-2-2-B}
s(r)\pp{}{r}
\longrightarrow
\pp{}{y}
\ ,
\end{eqnarray}
such that $\op{c}$ amounts to a shift of $-1$ in the coordinate
$y$. The transformations between $y$ and $r$ are denoted by
$Y(r)$ and $R(y)$
\begin{eqnarray}
r
&\stackrel{Y}{\longrightarrow}&
y=Y(r)
\\
y&\stackrel{R}{\longrightarrow}&
r=R(y)
\ .
\end{eqnarray}
They are by construction the inverse of each other
\begin{eqnarray}
R(Y(r)) = r
\qquad\mbox{and}\qquad
Y(R(y)) = y
\ .
\end{eqnarray}
From \fmref{E-2-2-B} follow the differential equations
\begin{eqnarray}\label{E-2-2-L}
\pp{}{r} Y(r) = \frac{1}{s(r)}
\qquad\mbox{and}\qquad
\pp{}{y} R(y) = s(R(y))
\end{eqnarray}
and the integral equation
\begin{eqnarray}\label{E-2-2-E}
y = \int^{R(y)} \frac{\dint\xi}{s(\xi)}
\ ,
\end{eqnarray}
which defines the transformation $R(y)$.
Rewriting eq. \fmref{E-2-2-C} in terms of $y$ and $r=R(y)$ one obtains
\begin{eqnarray}
\bra{R(y)\frac{\vek{r}}{r}}\op{c}\ket{\phi}
&=&
\frac{1}{R(y)\sqrt{s(R(y))}}\;
\exp\left\{-\pp{}{y}\!\!\right\}
\,R(y)\sqrt{s(R(y))}\;
\braket{R(y)\frac{\vek{r}}{r}}{\phi}
\nonumber
\\
\label{E-2-2-D}
&&\\
&=&
\frac{R(y-1)}{R(y)}
\sqrt{\frac{s(R(y-1))}{s(R(y))}}\;
\braket{R(y-1)\frac{\vek{r}}{r}}{\phi}
\nonumber
\ .
\end{eqnarray}
The action of $\exp\{-\pp{}{y}\}$ is to shift $y$ by $-1$ and
to leave the direction $\vecrdr$ unchanged.
Rewriting \fmref{E-2-2-D} in terms of $r$ yields
\begin{eqnarray}\label{E-2-2-F}
\bra{\vek{r}}\op{c}\ket{\phi}
&=&
\frac{R(Y(r)-1)}{r}
\sqrt{\frac{s(R(Y(r)-1))}{s(r)}}\;\;
\braket{R(Y(r)-1)\frac{\vek{r}}{r}}{\phi}
\end{eqnarray}
or for the inverse transformation
\begin{eqnarray}
\bra{\vek{r}}\op{c}^{-1}\ket{\phi}
&=&
\frac{R(Y(r)+1)}{r}
\sqrt{\frac{s(R(Y(r)+1))}{s(r)}}\;\;
\braket{R(Y(r)+1)\frac{\vek{r}}{r}}{\phi}
\ .
\end{eqnarray}
These two equations show that the transformations $R$ and $Y$
appear only in the form $R(Y(r)\pm 1)$. Therefore it is
convenient to introduce the functions
\begin{eqnarray}
R_{\pm}(r)
\equiv
R(Y(r)\pm 1)
\end{eqnarray}
with which we also express the terms under the square root as
\begin{eqnarray}
R_{\pm}^{\,\prime}(r)
\equiv
\pp{}{r}
R_{\pm}(r)
=
\frac{s(R_{\pm}(r))}{s(r)}
\ .
\end{eqnarray}
From the integral equation \fmref{E-2-2-E} follows
\begin{eqnarray}
\int_{R(y)}^{R(y\pm 1)}
\frac{\dint\xi}{s(\xi)}
=
y\pm1 - y
\qquad\mbox{or}\qquad
\int_{r}^{R_{\pm}}
\frac{\dint\xi}{s(\xi)}
=
\pm 1
\end{eqnarray}
which is the formal definition of $R_{\pm}(r)$.
From the fact that $Y(r)$ and $R(y)$ are inverse of each other follows
\begin{eqnarray}
R_{\pm}(R_{\mp}(r)) = r
\ ,
\end{eqnarray}
which means that $R_{+}$ is the inverse of $R_{-}$ and hence
\begin{eqnarray}\label{E-2-2-J}
R_{\pm}^{\,\prime}(r) = 
\left[R_{\mp}^{\,\prime}(R_{\pm}(r))\right]^{-1}
\ .
\end{eqnarray}
With help of the coordinate transformations $\RP{r}$ and
$\RM{r}$ the correlated wave function for the relative motion of
two particles given in \fmref{E-2-2-F} can be written as
\begin{eqnarray}\label{E-2-2-G}
\bra{\vek{r}}\op{c}\ket{\phi}
&=&
\frac{\RM{r}}{r}
\sqrt{\RMPr{r}}\;\;
\braket{\RM{r}\frac{\vek{r}}{r}}{\phi}
\end{eqnarray}
and the inverse transformation as
\begin{eqnarray}\label{E-2-2-H}
\bra{\vek{r}}\op{c}^{-1}\ket{\phi}
&=&
\frac{\RP{r}}{r}
\sqrt{\RPPr{r}}\;\;
\braket{\RP{r}\frac{\vek{r}}{r}}{\phi}
\ .
\end{eqnarray}
Hence, the unitary correlator $\op{c}=\exp\{-i\op{g}\}$ is
uniquely given by the coordinate transformation $\RP{r}$.

With the correlated state given in \fmref{E-2-2-G} one can
define the correlation volume $V_c$ by the square of the defect 
wave--function which is the difference between a uniform
uncorrelated state $\braket{\vek{r}}{\phi}$ and 
$\braket{\vek{r}}{\psi} = \bra{\vek{r}} \op{c} \ket{\phi}$
\begin{eqnarray}\label{E-2-2-P}
V_c 
:=&&
4 \pi \int_0^\infty \dint r \; \Big[r - \RM{r} \sqrt{\RMPr{r}} \Big]^2
\\
=&&
4 \pi \int_0^\infty \dint r \; \Big[r - \RP{r} \sqrt{\RPPr{r}} \Big]^2
\nonumber
\ .
\end{eqnarray}
A typical case is illustrated in \figref{F-2-2-2}.

By looking at a matrix element
$\bra{\psi_1}\op{b}(\vek{r})\ket{\psi_2}$ of a local operator
$\op{b}(\vek{r})$, which acts on the relative distance, the
meaning of the different parts in \fmref{E-2-2-G} and
\fmref{E-2-2-H} becomes clear
\begin{eqnarray}\label{E-2-2-I}
&&
\bra{\psi_1} \op{b}(\vek{r}) \ket{\psi_2}
=
\int r^2 \dint r \, \dint\Omega \;\;
\braket{\psi_1}{\vek{r}}\; b(\vek{r})\; \braket{\vek{r}}{\psi_2}
\\
&&\qquad =
\int r^2 \dint r \, \dint\Omega \;\;
\bra{\phi_1} \op{c}^{\dagger} \ket{\vek{r}}\; b(\vek{r})\;
\bra{\vek{r}}\op{c}\ket{\phi_2}
\nonumber
\\
&&\qquad =
\int r^2 \dint r \, \dint\Omega \;\;
\left(\frac{\RM{r}}{r}\right)^2 \RMPr{r}\;\;
\braket{\psi_1}{\RM{r}\frac{\vek{r}}{r}}\; b(\vek{r}) \;
\braket{\RM{r}\frac{\vek{r}}{r}}{\psi_2}
\nonumber
\\
&&\qquad =
\int r_{\!-}^2 \dint r_{\!-} \dint\Omega_{-} \;\;
\braket{\psi_1}{\vek{r}_{\!-}}\; 
b\left(\RP{r_{\!-}}\frac{\vek{r}_{\!-}}{r_{\!-}}\right) \;
\braket{\vek{r}_{\!-}}{\psi_2}
\nonumber
\ .
\end{eqnarray}
The factor $\frac{\RM{r}}{r}\sqrt{\RMPr{r}}$ in \fmref{E-2-2-G}
is just the square root of the Jacobian of the transformation
\begin{eqnarray}\label{E-2-2-K}
&&
\vek{r} \longrightarrow
\vek{r}_{\!-} \equiv
\RM{r} \frac{\vek{r}}{r}
\\[2mm]
&&
\dint^3 r_{\!-}
=
r_{\!-}^2 \dint r_{\!-} \dint\Omega_{-}
=
\left(\frac{\RM{r}}{r}\right)^2 \RMPr{r}\;
r^2 \dint r \, \dint\Omega
\ ,
\end{eqnarray}
where
$\dint\Omega=\dint\Omega_{-}=\sin\theta\dint\theta\dint\phi$
remains unchanged.

Thus, the correlated wave function \fmref{E-2-2-G} at a relative
distance $\vek{r}$ is given by the uncorrelated one taken at a
closer distance $\RM{r}\vecrdr$ times the square root of the
Jacobian which takes care of unitarity.

Equivalently one can use the correlated relative distance
\begin{eqnarray}
\coop{\vek{r}} = \op{c}^{-1}\;\vek{r}\;\op{c}
=
\RP{r}\frac{\vek{r}}{r}
\ .
\end{eqnarray}
Due to the unitarity property $\op{c}^{\dagger}\op{c}=1$ any
local operator $\op{b}(\vek{r})$ can be readily transformed to
the correlated one
\begin{eqnarray}\label{E-2-2-M}
\coop{b(\vek{r})} = \op{c}^{\dagger}\;\op{b}(\vek{r})\;\op{c}
=
\op{c}^{-1}\;\op{b}(\vek{r})\;\op{c}
=
b\left(\RP{r}\frac{\vek{r}}{r}\right)
\ .
\end{eqnarray}
The correlated operator $\coop{b(\vek{r})}$ is just the
uncorrelated one taken at the transformed relative distance
$\RP{r}\vecrdr$. This is also the physical meaning of the last
line in eq. \fmref{E-2-2-I}.

The correlated radial momentum $\coop{q}_r$ can be calculated
in analogy to \fmref{E-2-2-I} by using the coordinate
transformation \fmref{E-2-2-K} $\vek{r}\rightarrow\vek{r}_{\!-}$
and the relation \fmref{E-2-2-J} between $R_{-}^\prime$ and
$R_{+}^\prime$. The result is
\begin{eqnarray}\label{E-2-2-O}
\coop{q}_r = \op{c}^{-1}\,q_r\;\op{c}
=
\frac{1}{\sqrt{\RPPr{r}}}\;
\frac{1}{r}\; q_r\; r\; \frac{1}{\sqrt{\RPPr{r}}}
\end{eqnarray}
where
\begin{eqnarray}
\bra{\vek{r}} q_r \ket{\phi} 
=
\bra{\vek{r}} \frac{\vek{r}}{r}\cdot\vek{q} \ket{\phi}
=
-i\, \pp{}{r}\,\braket{\vek{r}}{\phi}
\ .
\end{eqnarray}
The correlated relative angular momentum is the same as the
uncorrelated one
\begin{eqnarray}\label{E-2-2-N}
\coop{\vek{l}} = \op{c}^{-1}\,\op{\vek{l}}\;\op{c}
=
\op{c}^{-1}\,\vek{r}\times\vek{q}\;\op{c}
=
\op{\vek{l}}
\ .
\end{eqnarray}
The present correlator conserves angular momentum and spin for
fermions. The realistic nucleon--nucleon interaction, however,
induces strong tensor correlations. For such a situation, which
will be subject of a forthcoming paper, the coordinate
transformation
$\vek{R}_{-}(\vek{r},\vek{\sigma}_1,\vek{\sigma}_2)$ will not
only depend on the relative distance $|\vek{r}|$ but also on the
angles between $\vek{r}$ and the spins $\vek{\sigma}_1$ and
$\vek{\sigma}_2$ of the two particles.

An intuitive relation between the coordinate transformation
$\RPM{r}$ and the original $s(r)$ in eq. \fmref{E-2-2-A} can be
obtained from the first term of the Taylor expansion around $y$
using \fmref{E-2-2-L}
\begin{eqnarray}
R_{\pm}(r)
&=&
R(Y(r)\pm 1)
=
R(Y(r)) \pm R^{\prime}(Y(r)) +\cdots
\\
&=&
r \pm s(r) +\cdots
\nonumber
\ .
\end{eqnarray}
For $s(r)$ small compared to $r$, $s(r)$ is the amount by which the two
particles are shifted away from their original distance $r$, as
anticipated earlier.

Instead of discussing the shift function $s(r)$ as the
variational degree of freedom we will consider the equivalent
coordinate transformation $\RPM{r}$ directly.

\begin{figure}[!bt]
\begin{center}
\epsfig{file=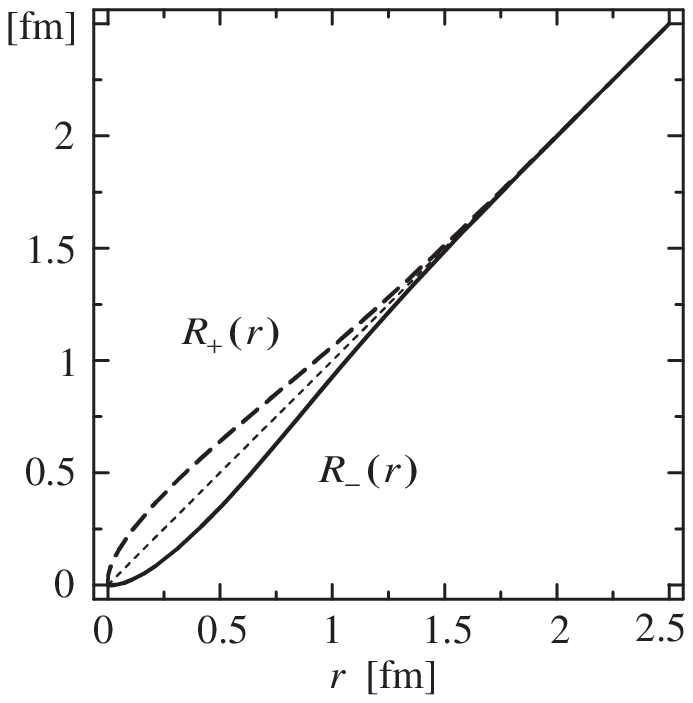,height=65mm}
\hspace{1ex}
\epsfig{file=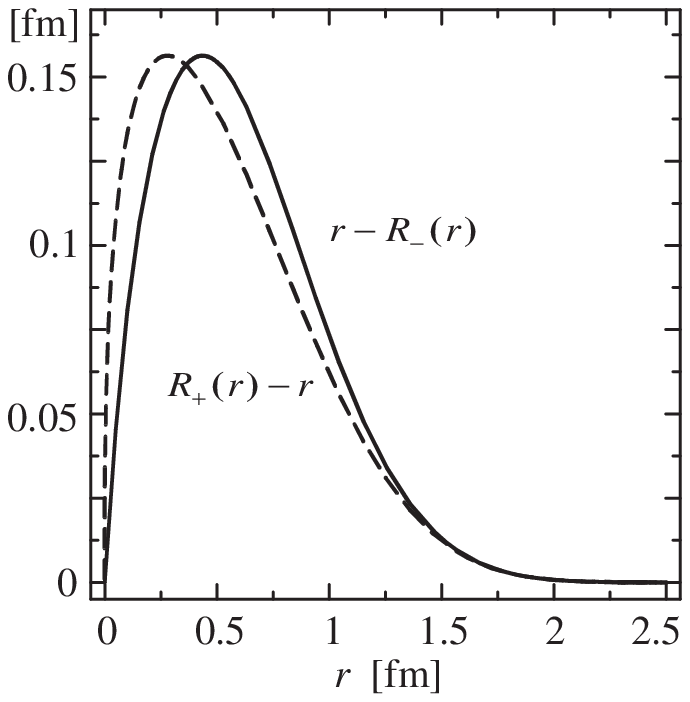,height=65mm}
\end{center}
\mycaption{L.h.s.: typical coordinate transformation $\RP{r}$ 
and its inverse $\RM{r}$; r.h.s: shifts $r-\RM{r}$ (full line)  
and $\RP{r}-r$ (dashed line).}{F-2-2-1}
\end{figure} 
\begin{figure}[!bt]
\vspace*{8mm}
\begin{center}
\epsfig{file=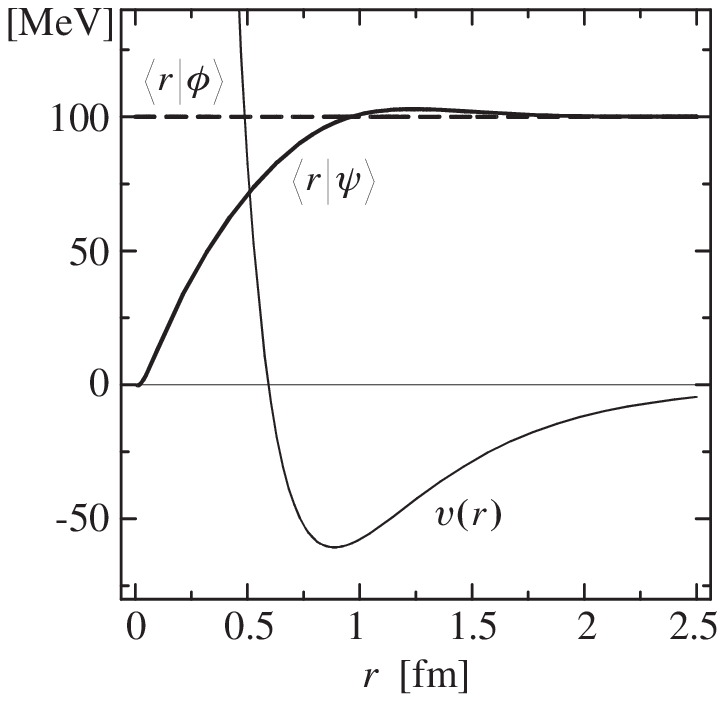,height=65mm}
\end{center}
\mycaption{Uncorrelated state $\braket{r}{\phi}$ (dashed line),
correlated state $\braket{r}{\psi}$ (full line) and potential
$v(r)$ (thin line) as a function of the relative distance $r$.}{F-2-2-2}
\end{figure} 

In order to avoid unphysical transformations, $\RPM{r}$ has to
increase monotonically with $r$ and has to be differentiable
everywhere. For correlations of finite range 
$\RPM{r}\rightarrow r$ for large $r$.
A typical example which will be used later is shown in
\figref{F-2-2-1} in terms of $\RPM{r}$ on the l.h.s.. The r.h.s.
displays the shifts $r-\RM{r}$ and $\RP{r}-r$ which is the
essential information. The shifts rise rapidly at $r=0$ to a
maximum of $0.15\fm$ around $r=0.4\fm$. This results in a strong
depletion of the relative wave function as can be seen in
\figref{F-2-2-2} where a uniform uncorrelated state is
compared with the correlated one as given in
eq. \fmref{E-2-2-G}. The corresponding potential is also
included in the figure. 
\subsection{One-- and two--body part of the correlated Hamiltonian}

Of particular interest is the $n$--body expansion of the
correlated Hamiltonian. Starting from eq. \fmref{E-2-2-M} the
two--body part of the correlated potential is given by
\begin{eqnarray}
\coop{v}^{[2]}(i,j)
=
\op{c}(i,j)^{-1} \;
\op{v}^{[2]}(i,j) \;
\op{c}(i,j) 
=
\op{v}\left(\RP{r_{ij}}\right)
\end{eqnarray}
where $\vek{r}_{ij}=\vek{x}(i)-\vek{x}(j)$ and
${r}_{ij}=|\vek{r}_{ij}|$. 

The three-body part $\coop{v}^{[3]}(i,j,k)$ assumes the more
complicated form given in \fmref{E-2-1-D}. It will be discussed
in section \xref{Sec-3-3}.

The correlated kinetic energy contains the uncorrelated
kinetic energy as the first term
\begin{eqnarray}
\coop{t}^{[1]}(i)
=
\op{t}(i)
\end{eqnarray}
because $\op{g}^{[2]}(i,j)$ is a two--body operator and hence
$\op{C}_1=1$ in the one--body space. The two--body part can be
calculated by first separating the kinetic energy of two
particles into relative and centre of mass energy
\begin{eqnarray}
\op{t}(i) + \op{t}(j)
=
\frac{1}{m} \; \vek{q}_{ij}^2
+
\frac{1}{4 m} \left(\vek{p}(i) + \vek{p}(j)\right)^2
\end{eqnarray}
where $\vek{q}_{ij}=(\vek{p}(i)-\vek{p}(j))/2$ is the
relative momentum.  Since $\op{g}^{[2]}(i,j)$ commutes with the centre of
mass coordinate, only the relative momentum is correlated
\begin{eqnarray}
&&\op{c}(i,j)^{-1} 
\left(\op{t}(i) + \op{t}(j)\right) \;
\op{c}(i,j)
\\
&&\qquad
=
\frac{1}{m} \; 
\op{c}(i,j)^{-1} \;
\vek{q}_{ij}^2 \;
\op{c}(i,j)
+
\frac{1}{4 m} \left(\vek{p}(i) + \vek{p}(j)\right)^2
\nonumber
\ .
\end{eqnarray}
Decomposing $\vek{q}_{ij}^{\,2}$ into a radial and angular momentum part
\begin{eqnarray}
\vek{q}_{ij}^{\,2}
=
q_{r\,ij}^\dagger \, q_{r\,ij}\;
+\;
\frac{1}{r_{ij}^2} \vek{l}_{ij}^2
\ ,
\end{eqnarray}
where $q_{r}=\frac{\vek{r}}{r}\cdot\vek{q}$ and 
$\vek{l}=\vek{r}\times\vek{q}$,
one can utilize eqs. \fmref{E-2-2-O} and \fmref{E-2-2-N} to obtain
\begin{eqnarray}
&&\op{c}(i,j)^{-1} \, 
\vek{q}_{ij}^{\,2} \;
\op{c}(i,j)
\\
&&\qquad
=
\frac{r_{ij}}{\sqrt{\RPPr{r_{ij}}}}\;
q_{r\,ij}^\dagger\;
\frac{1}{r_{ij}^2 \RPPr{r_{ij}}}\;
q_{r\,ij}\;
\frac{r_{ij}}{\sqrt{\RPPr{r_{ij}}}}\;
+
\frac{1}{(\RP{r_{ij}})^2}
\op{\vek{l}}_{ij}^{\,2}
\nonumber
\ .
\end{eqnarray}
Commuting the two radial momenta $q_{r\,ij}$ in the first term
on the r.h.s. to the left and to the right of
$r_{ij}/\sqrt{\DRP{r_{ij}}}$, respectively, yields
\begin{eqnarray}\label{E-2-3-AA}
&&\op{c}(i,j)^{-1} \,
\vek{q}_{ij}^{\,2} \;
\op{c}(i,j)
\\
&&\qquad
= 
q_{r\,ij}^\dagger\;
\frac{1}{(\DRP{r_{ij}})^2}\;
q_{r\,ij}\
+
\frac{1}{(\RP{r_{ij}})^2}
\op{\vek{l}}_{ij}^{\,2}
+
m\; \coop{u}^{[2]}(i,j)
\nonumber
\ ,
\end{eqnarray}
where
\begin{eqnarray}\label{E-2-3-A}
\!\!\coop{u}^{[2]}(i,j)
=
\frac{1/m}{(\DRP{r_{ij}})^2}
\left\{
2\frac{\DDRP{r_{ij}}}{r_{ij} \DRP{r_{ij}}}
-\frac{5}{4}
\left(\frac{\DDRP{r_{ij}}}{\DRP{r_{ij}}}\right)^2
+\half\frac{\DDDRP{r_{ij}}}{\DRP{r_{ij}}}
\right\}
\ .
\end{eqnarray}
$\DDRP{r}$ and $\DDDRP{r}$ are the second and the third
derivative of $\RP{r}$ with respect to $r$.  Thus, the two--body
part of the correlated kinetic energy may be split into three
parts
\begin{eqnarray}\label{E-2-3-C}
&&\coop{t}^{[2]}(i,j)
\\
&&=
\op{q}_{r\,ij}^\dagger\;
\frac{1}{m} \left[
\frac{1}{(\DRP{r_{ij}})^2} - 1 \right]
\op{q}_{r\,ij}
+
\frac{1}{m} \left[
\frac{1}{(\RP{r_{ij}})^2} - \frac{1}{r_{ij}^2}\right]
\op{\vek{l}}_{ij}^{\,2}
+
\coop{u}^{[2]}(i,j)
\nonumber
\ .
\end{eqnarray}
The first is a two--body radial kinetic energy with a reduced
radial mass
\begin{eqnarray}\label{E-2-3-D}
\mu_r(r)
=
\frac{m}{2} \left[
\frac{1}{(\DRP{r})^2} - 1 \right]^{-1}
\end{eqnarray}
which depends on the distance $r$ between the particles in
the pair.
The second part is a relative rotational energy with a reduced
angular mass
\begin{eqnarray}\label{E-2-3-E}
\mu(r)
=
\frac{m}{2} \left[
\frac{r^2}{(\RP{r})^2} - 1 \right]^{-1}
\ .
\end{eqnarray}
The third part $\coop{u}^{[2]}(i,j)$ is a local two--body
potential which depends on derivatives of $\RP{r}$ as given
by eq. \fmref{E-2-3-A}. 
All three parts vanish outside the correlation range when
$\RP{r}\rightarrow r$.

The one-- and two--body parts of the correlated Hamiltonian add up
to
\begin{eqnarray}\label{E-2-3-B}
\coop{H}^{[1]} + \coop{H}^{[2]}
&&=
\sum_{i}^A\; \op{t}(i)
\\
&&+
\sum_{i<j}^{A} \left[
\op{q}_{r\,ij}^\dagger\;
\frac{1}{2 \mu_r(r_{ij})}\;
\op{q}_{r\,ij}
+
\frac{1}{2 \mu(r_{ij}) r_{ij}^2}\;
\op{\vek{l}}_{ij}^2
\right]
\nonumber
\\
&&+
\sum_{i<j}^{A} \left[
\coop{u}^{[2]}(i,j)
+
\coop{v}^{[2]}(i,j)
\right]
\nonumber
\ .
\end{eqnarray}
The first part is the uncorrelated one--body kinetic energy. The
second is a momentum dependent two--body interaction which is
restricted to the correlation area. The third is the tamed
two--body potential which has no repulsive core anymore. Its
shape will be determined in the following section where the
choice of $\RP{r}$ is discussed. The taming of $\op{v}(r)$ has
to be paid by the momentum dependent two--body interaction. In
Cartesian coordinates this part may also be written as
\begin{eqnarray}
\op{q}_{r}^\dagger\;
\frac{1}{2 \mu_r(r)}\;
\op{q}_{r}
+
\frac{1}{2 \mu(r) r^2}\;
\op{\vek{l}}^2
=
\op{\vek{q}}
\frac{1}{2 \mu(r)}\;
\op{\vek{q}}
+
\op{q}_{r}^\dagger\;
\frac{1}{2 \mu_r^*(r)}\;
\op{q}_{r}
\ ,
\end{eqnarray}
where
\begin{eqnarray}
\mu_r^*(r)
=
\frac{m}{2}
\left[\frac{1}{(\DRP{r})^2} - \frac{r^2}{(\RP{r})^2}\right]^{-1}
\ .
\end{eqnarray}

\subsection{Choice of the correlation function $R_{\pm}(r)$}
\label{Sec-2-4}

The correlator is a functional of $s(r)$ or equivalently of the
coordinate transformation $\RM{r}$
\begin{eqnarray}
\ket{\Psi} = \op{C}[R_{-}]\;\ket{\Phi}
\ .
\end{eqnarray}
Therefore, $\RM{r}$ represents additional variational freedom
and according to the Ritz variational principle $\RM{r}$ could
be determined by minimizing the ground state energy
\begin{eqnarray}
\bra{\Psi} \op{H} \ket{\Psi}
=
\bra{\Phi}\op{C}^\dagger[R_{-}]\;\op{H}\;\op{C}[R_{-}]\;\ket{\Phi}
\end{eqnarray}
not only with respect to $\ket{\Phi}$ but also with respect to
$\RM{r}$.

However, since only the first terms in the $n$--body expansion of the
correlated Hamiltonian $\coop{H} = \coop{H}^{[1]} +
\coop{H}^{[2]} + \cdots$ (see eq. \fmref{E-2-1-A})
will be used, the exact ground state energy is in general no
lower bound on
$\bra{\Phi}\coop{H}^{[1]}\ket{\Phi}+\bra{\Phi}\coop{H}^{[2]}\ket{\Phi}$
anymore because contributions
$\bra{\Phi}\coop{H}^{[n]}\ket{\Phi}$ for $n\ge3$ can be positive
and negative so that only the complete sum up to the total
particle number $A$ has the lower bound.  

In order to make the three-- and higher--body correlation part
as small as possible one should use the freedom in the
separation of short and long range correlations, 
i.e. what should be taken care of by $\op{C}[R_{-}]$ 
and $\ket{\Phi}$ respectively, to
find an optimal correlated two--body Hamiltonian.  One expects
the three--body part of the correlated Hamiltonian which is
induced by $\op{C}[R_{-}]$ to be small if the correlation range
in which $\RM{r}\ne r $ (or $s(r)\ne 0$) is short compared to
the mean distance between the particles. A measure of the
smallness is the uncorrelated density $\rho_0$ times the
correlation volume $V_{c}$, defined in \fmref{E-2-2-P}. 
In section \xref{Sec-3-3} where we calculate three--body terms a
relation between $\rho_0 V_c$ and the three--body correlations
is given. From that one can judge how small $\rho_0 V_c$
actually should be.

In the following we investigate for a two--body system the
general idea that the correlator $\op{C}$ can be imagined as a
unitary mapping of uncorrelated states onto correlated states
which approximately diagonalize the Hamiltonian $\op{H}$.
\begin{eqnarray}
\op{H} \ket{\psi_E}
=
E \ket{\psi_E}
\end{eqnarray}
$\ket{\psi_E}$ are the exact bound and scattering eigenstates of
the Hamiltonian $\op{H}=\frac{1}{2\mu}\op{\vek{q}}^2+\op{v}(r)$
for the relative motion. The repulsive core of $\op{v}(r)$ will
suppress the wave function $\braket{\vek{r}}{\psi_E}$ for small
$r$. The goal is to find for a range of energies an optimal
correlator $\op{c}[R_{-}]$ which removes these repulsive
correlations from $\ket{\psi_E}$.

If one demands that in a range $r\le\lambda$ the correlated
trial state and the exact state coincide
\begin{eqnarray}\label{E-2-4-G}
\bra{\vek{r}} \op{c}[R_{-}]\;\ket{\phi}
\stackrel{!}{=}
\braket{\vek{r}}{\psi_E}
\quad\mbox{for}\quad 0\le |\vec{r}| \le\lambda
\end{eqnarray}
one obtains for angular momentum equal zero from eq. \fmref{E-2-2-G}
\begin{eqnarray}\label{E-2-4-D}
\sqrt{\DRM{r}} \; u(\RM{r}) = u_E(r)
\end{eqnarray}
or the integral equation
\begin{eqnarray}\label{E-2-4-F}
\RM{r}
=
\int_{0}^{r} \dint\xi
\left(
\frac{u_E(\xi)}{u(\RM{\xi})}
\right)^2
\ ,
\end{eqnarray}
where $\RM{r=0}=0$ is assumed and the radial wave functions are
denoted as
\begin{eqnarray}
u_E(r) = r \braket{\vek{r}}{\psi_E}
\quad\mbox{and}\quad
u(r) = r \braket{\vek{r}}{\phi}
\ .
\end{eqnarray}
As already discussed in a more general frame in section
\xref{Sec-2-0} the correlator $\op{C}$ or the coordinate
transformation $\RM{r}$ as defined in eq. \fmref{E-2-4-F} depend
on the trial function $u(r)$ and on the Hamiltonian via its
eigenfunctions $u_E(r)$.

For systems at small energies a first attempt is to assume the
trial state $\braket{\vek{r}}{\phi}=1/\sqrt{N_0}$ to be uniform
across the repulsive core \cite{PaW79}.  For this
special case the coordinate transformation $\RM{r}$ is given by
the differential equation
\begin{eqnarray}\label{E-2-4-H}
\DRM{r}
=
N_0 \left(\frac{u_E(r)}{\RM{r}}\right)^2
\ ,
\end{eqnarray}
where the norm $N_0$ is still to be chosen. 

\begin{figure}[!bt]
\begin{center}
\epsfig{file=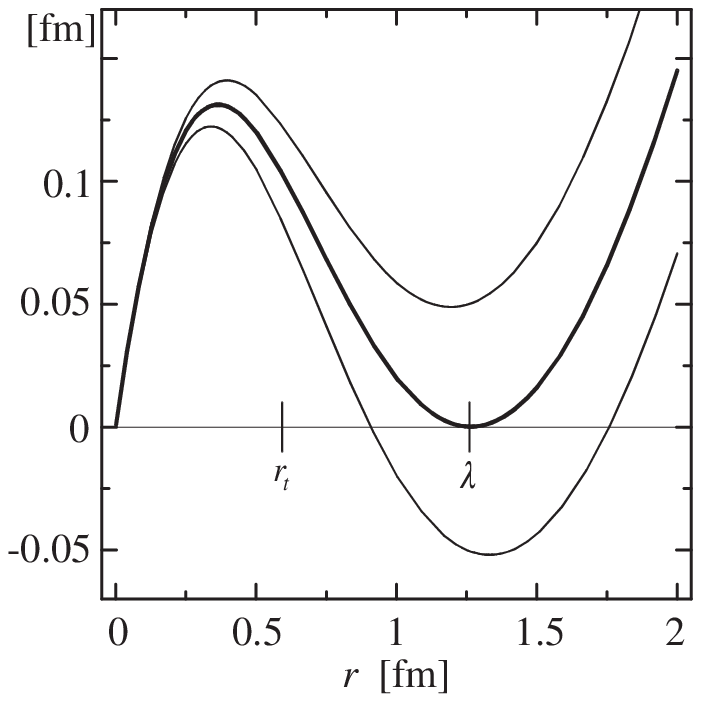,height=65mm}
\hspace{1ex}
\epsfig{file=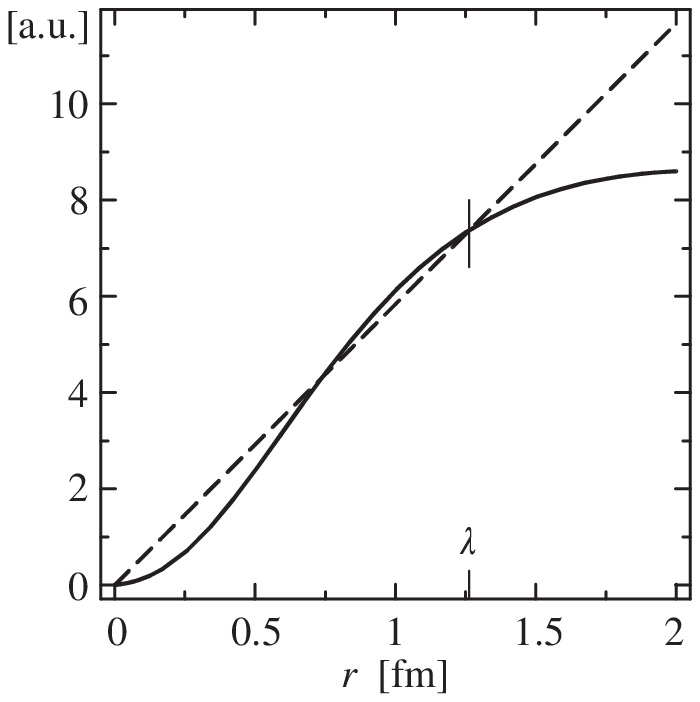,height=65mm}
\end{center}
\mycaption{L.h.s.: Coordinate transformation $\RM{r}$ plotted as
$r-\RM{r}$ for $E=0$ and different $N_0$.
R.h.s: Exact radial wave function $u_{E=0}(r)$ (solid line) 
and trial function $u(r)$ for which $\RM{\lambda}=\lambda$ and 
$\DRM{\lambda}=1$.}{F-2-4-2}
\end{figure} 

The dependence of $\RM{r}$ on $N_0$, which is displayed in
\figref{F-2-4-2} l.h.s. in terms of the deviation $r-\RM{r}$,
will now be used to limit the range of the correlation.  For
this the MTV potential (\cite{MaT69} and eq. \fmref{E-A-A}),
already shown in \figref{F-2-2-1}, is chosen as an example and
the energy $E$ is set to zero. Among the set of curves there is
one which touches the $x$--axis at a distance $\lambda$ which
means
\begin{eqnarray}\label{E-2-4-A}
\RM{\lambda} = \lambda
\quad\mbox{and}\quad
\DRM{\lambda} = 1
\ .
\end{eqnarray}
From the Schr\"odinger equation for angular momentum zero
\begin{eqnarray}\label{E-2-4-I}
u_E^{\prime\prime}(r)
=
m \; (\op{v}(r) - E) \; u_E(r)
\end{eqnarray}
one sees that at the classical turning point $v(r_t)=E$ the
curvature of $u_E(r)$ switches sign so that a linear radial
function $u(r)=r/\sqrt{N_0}$ can be found which intersects the
exact solution twice, see \figref{F-2-4-2} r.h.s., such that at
$r=\lambda$ trial and exact state coincide
\begin{eqnarray}\label{E-2-4-B}
u(\lambda) = u_E(\lambda)
\end{eqnarray}
and their norms up to $\lambda$ are also equal
\begin{eqnarray}\label{E-2-4-C}
\int_0^\lambda \dint r \; (u(r))^2
=
\int_0^\lambda \dint r \; (u_E(r))^2
\ .
\end{eqnarray}
The transformation $\RM{r}$ which fulfills the conditions
\fmref{E-2-4-A} or equivalently \fmref{E-2-4-B} and
\fmref{E-2-4-C} provides a natural division into a short and a
long range part. 
This division is an inherent property of all methods which
derive effective interactions \cite{MoS60}.
At $r=\lambda$ one may split $\op{c}[R_{-}]$ into
\begin{eqnarray}
\op{c}[R_{-}]
=
\op{c}[R_{-}^{I}] \; \op{c}[R_{-}^{II}]
\end{eqnarray}
where
\begin{eqnarray}
R_{-}^{I}
=
\left\{
\begin{array}{c@{\;\mbox{for}\;}c}
\RM{r} & 0 \le r \le \lambda \\
r      & \lambda \le r
\end{array}\right.
\quad\mbox{and}\quad
R_{-}^{II}
=
\left\{
\begin{array}{c@{\;\mbox{for}\;}c}
r      & 0 \le r \le \lambda \\
\RM{r} & \lambda \le r
\end{array}\right.
\end{eqnarray}
because $\op{c}[R_{-}^{I}]$ commutes with $\op{c}[R_{-}^{II}]$ if
\fmref{E-2-4-A} is fulfilled. For $r>\lambda$ the uncorrelated
state $\braket{\vek{r}}{\phi}$ is supposed to describe the long
range part so that we set $R_{-}^{II}=r$ or
$\op{c}[R_{-}^{II}]=1$.

The schematic case of a uniform uncorrelated state illustrates
nicely how the method works. As the correlation function
$\RM{r}$ is chosen such that the correlated state equals the
exact solution for a given energy $E$ (see eq. \fmref{E-2-4-G})
it also solves the Schr\"odinger equation \fmref{E-2-4-I}
\begin{eqnarray}
\bra{\vek{r}}
\left( \frac{1}{m} \op{q}^\dagger_r\, \op{q}_r + v(r) \right)
 \op{c}[R_{-}]\;\ket{\phi}
=
E \; \bra{\vec{r}} \; \op{c}[R_{-}]\; \ket{\phi}
\end{eqnarray}
or
\begin{eqnarray}
\bra{\vek{r}}
\left( \frac{1}{m} 
\op{c}[R_{-}]^{-1} \; \op{q}^\dagger_r \, \op{q}_r \; \op{c}[R_{-}]\; 
+ \op{c}[R_{-}]^{-1} \; v(r) \; \op{c}[R_{-}] \right)
\ket{\phi}
=
E \; \braket{\vec{r}}{\phi}
\ .
\end{eqnarray}
Since $\braket{\vek{r}}{\phi}=1/\sqrt{N_0}$ is uniform, the
relative kinetic energy reduces to (c.f. eqs. \fmref{E-2-3-AA}
and \fmref{E-2-3-A})
\begin{eqnarray}\label{E-2-4-J}
\bra{\vek{r}}
\frac{1}{m} 
\op{c}[R_{-}]^{-1} \; \op{q}^\dagger_r\, \op{q}_r\; \op{c}[R_{-}]\; 
\ket{\phi}
=
\coop{u}^{[2]}(r) \frac{1}{\sqrt{N_0}}
\end{eqnarray}
and hence
\begin{eqnarray}\label{E-2-4-K}
\coop{u}^{[2]}(r) + \coop{v}^{[2]}(r) = E
\quad\mbox{for}\quad r \le\lambda
\ .
\end{eqnarray}
This means that a correlator obtained from eq. \fmref{E-2-4-H}
for $E=0$ transforms the original Hamiltonian for small
distances $r<\lambda$, where the potential is strongly
repulsive, into a correlated Hamiltonian with a vanishing
local potential $\coop{u}^{[2]}(r)+\coop{v}^{[2]}(r)=0$. 
Thus the repulsive core is completely absent in the 
correlated Hamiltonian.

If the potential is purely repulsive ($v(r)\ge0$ for all $r$) 
there is no natural scale for the division into long and short 
ranged components. 
In this case the quest for a correlation volume small enough for
the two--body approximation will determine the range of the 
correlation.
Further discussions on repulsive potentials can be found in section
\xref{Sec-4-1} where we treat spin--isospin--dependent correlators. 

\begin{figure}[!bt]
\begin{center}
\epsfig{file=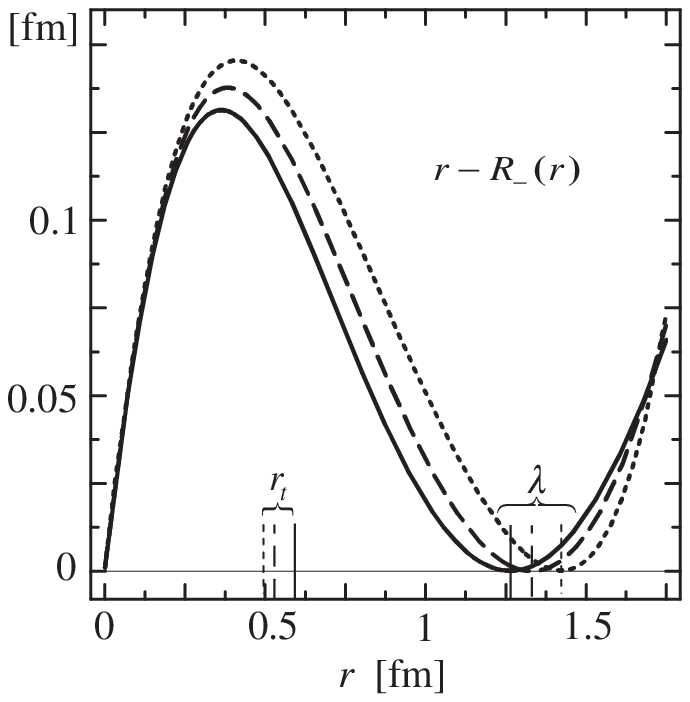,height=65mm}
\end{center}
\mycaption{Energy dependence of the coordinate transformation 
$\RM{r}$, $E=0$ (full line), $E=40\MeV$ (dashed line),
$E=80\MeV$ (dotted line). $r_t$ classical turning points,
$\lambda$ range of correlators.}{F-2-4-3}
\end{figure} 

The next point is the dependence on the energy.
Fig. \xref{F-2-4-3} displays $r-\RM{r}$ for a separable
correlation at different energies between $0$ and $80$MeV which
is the kinetic energy of two nucleons colliding with opposite
Fermi momenta. As the uncorrelated state we choose the spherical
Bessel function so that the radial function
$u(r)=rj_0(\sqrt{mE}r)/\sqrt{N_0}=\sin(\sqrt{mE}r)/\sqrt{N_0}$
is the $l=0$ solution of the free Schr\"odinger equation.
It turns out that for distances $r<r_t$ the correlation 
function $\RM{r}$ is rather energy independent. For
$r>r_t$ one expects an influence from the uncorrelated state so that
the shape of $\RM{r}$ may differ for $r_t<r<\lambda$ from the
one shown in \figref{F-2-4-3} depending on the choice of the
uncorrelated state.

The kink in the wave function at $r=\lambda$ (see
\figref{F-2-4-2} r.h.s.) caused by cutting off $\RM{r}$ at $r=\lambda$ in a
non-analytic way should be removed. Therefore, the short range
part $R_{-}^{I}(r)$ will be parameterized by an analytic function
which approximates $\RM{r}$ for $0<r<r_t$ rather well and leaves
some freedom for $r>r_t$ for adjusting to the actual trial
states used.

Since later on in the many--body case the product state
will consist of Gaussian shaped single--particle states, the
uncorrelated state of the two--particle system, from which the
coordinate transformation $\RM{r}$ is deduced, is also assumed to
be a Gaussian at small $r$ (and not a constant or a Bessel
function) which joins smoothly to an exponential radial 
function at $r=\rho$
\begin{eqnarray}\label{E-2-4-E}
\braket{\vek{r}}{\phi}
=
\frac{1}{\sqrt{N(\rho)}}
\left\{
\begin{array}{c@{\quad\mbox{for}\quad}c}
\exp\left\{-\frac{\rho+\kappa}{2\rho^2 \kappa} r^2 \right\} & r \le \rho\\
\frac{\rho}{r}\exp\left\{\frac{\rho-\kappa}{2 \kappa}-\frac{r}{\kappa}\right\}
& r \ge \rho
\end{array}\right.
\ .
\end{eqnarray}
The length $\kappa=(-m E_0)^{-1/2}$ is chosen such that the
long range part, where the correlator should not modify the
trial state, has the correct fall off.

\begin{figure}[!bt]
\begin{center}
\epsfig{file=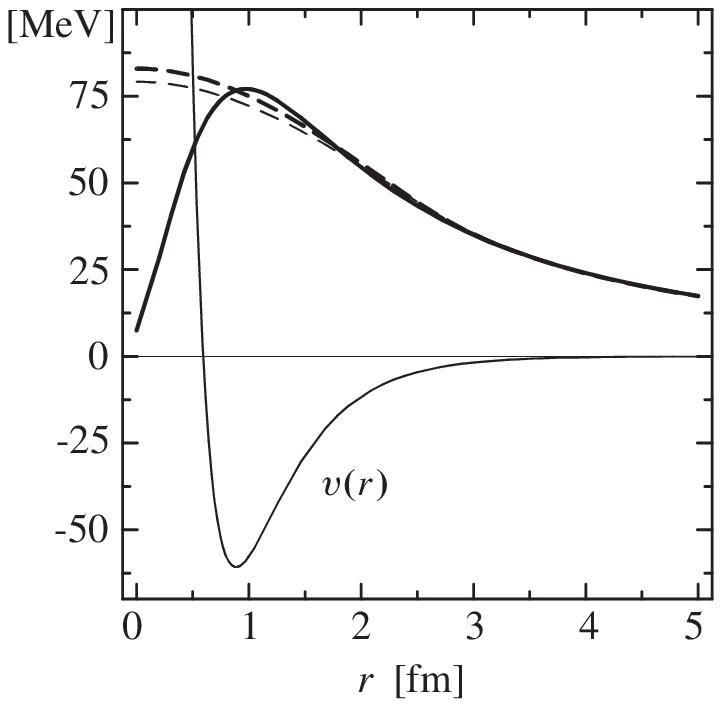,height=65mm}
\hspace{1ex}
\epsfig{file=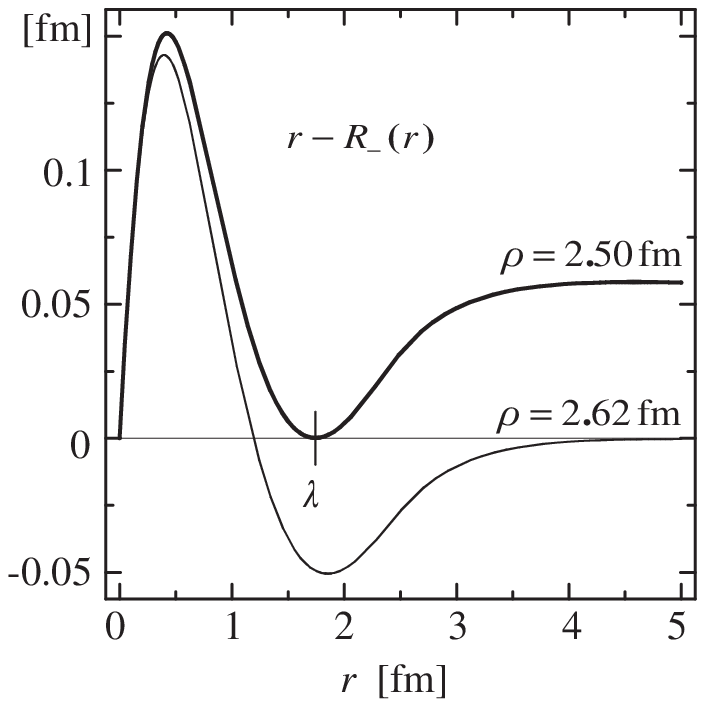,height=65mm}
\end{center}
\mycaption{L.h.s.: MTV potential $v(r)$ in MeV, uncorrelated wave
functions for $\rho=2.62\fm$ (thin dashed line) and
$\rho=2.5\fm$ (thick dashed line) and corresponding correlated
ones which both equal the exact one (thick solid line).  R.h.s:
correlation functions in terms of $r-\RM{r}$ for uncorrelated
states which map on the exact ground state.}{F-2-4-5}
\end{figure} 

Fig. \xref{F-2-4-5} displays on the left hand side the
uncorrelated trial state $\braket{\vek{r}}{\phi}$ for
$\rho=2.62\fm$ and the exact ground state
$\braket{\vek{r}}{\psi_E}$ together with the MTV potential. The
exact solution is pushed out of the repulsive region compared to
the uncorrelated state. The coordinate transformation $\RM{r}$
obtained from \fmref{E-2-4-F}, which transforms
$\braket{\vek{r}}{\phi}$ into $\braket{\vek{r}}{\psi_E}$, is
shown on the r.h.s. in terms of the shift distance $r-\RM{r}$ 
(thin line). Due to the attraction this shift turns out to be
negative beyond $r=1.2\fm$ so that more probability is
accumulated around $r\approx1\fm$ where the potential is most
attractive. This long range effect should not be corrected by
the correlator but an improved trial state should be employed,
otherwise the three-- and higher--body terms in the correlated
Hamiltonian will be too large.

Similar to the uniform trial state discussed above we vary the
norm $N(\rho)$ by moving the matching point $\rho$ until we
achieve a separable correlator which is the case for
$\rho=2.5\fm$ (thick line in \figref{F-2-4-5} r.h.s.). This
correlation function differs only little from the one where
trial and exact state coincide for very large $r$
($\rho=2.62\fm$).  The resulting shifts $r-\RM{r}$ are displayed
on the r.h.s. of \figref{F-2-4-5}. For $r<r_t=0.6\fm$ they are
identical and similar to the case of a uniform uncorrelated
state. The separation point $\lambda$ turns out to be inside
$r = \rho$, so that the correlation acts only on the Gaussian part.

\section{Few--body systems with spin--isospin--independent forces}
\label{Sec-3-0}

In this section we apply the concept of the unitary
correlation operator to the two--, three-- and four--body
systems \element{2}{H} and \element{3,4}{He}. The correlation
function $\RP{r}$ is first parameterized and then the binding
energies of the three nuclei are minimized with respect to the
parameters of $\RP{r}$. It turns out that the resulting
correlation functions are almost identical. This shows that the
general concept of a state--independent correlator discussed in
the previous section is applicable. The last subsection
investigates an approximate treatment of the three--body term
in the correlated Hamiltonian.

Throughout this section the Malfliet--Tjon V potential (MTV)
\cite{MaT69} is used, which is a central interaction without spin and
isospin dependence (see appendix eq. \fmref{E-A-A}).
This potential is widely used in the literature, for instance to
compare different theoretical many--body approaches
\cite{VaS95,HeZ85} because it has a repulsive core and an 
attractive long range part chosen by considering the phase 
shifts of the scattering problem. 
However, already for \element{4}{He} it overbinds. 
For larger nuclei it gives completely unrealistic
results. Therefore another potential will be used in section
\xref{Sec-4-0} where nuclei up to $A=48$ are investigated.

\subsection{Parameterization of $\RPM{r}$}

For the two--body system everything can be calculated with a
correlation function $\RP{r}$ that is derived as described
before and at least numerically known at all $r$. But for
many--body systems it is advisable to have an analytic form with
a limited number of parameters. Therefore we parameterize the
numerically determined $\RP{r}$ by
\begin{eqnarray}\label{E-3-1-A}
\RP{r}
=
r + \alpha \left(\frac{r}{\beta}\right)^{\eta}
\exp\!\left\{-\exp\!\left(\frac{r}{\beta} \right)   \right\}
\ .
\end{eqnarray}
The parameter $\alpha$ controls the overall amount of the shift
and $\beta$ the length scale. $\eta$ determines the steepness around
$r=0$. The double--exponential fall off is required to reproduce
the exact correlation function in the vicinity of the separation
distance $r=\lambda$, which has been derived in the previous
section.  Figure \xref{F-3-1-1} compares the exact correlation
function for the bound state with the parameterization \fmref{E-3-1-A},
where $\alpha=0.935\fm$, $\beta=0.95\fm$ and $\eta=0.39$. A
similar fit can be achieved with a Gaussian replacing the
double--exponential fall off.

\begin{figure}[!bt]
\begin{center}
\epsfig{file=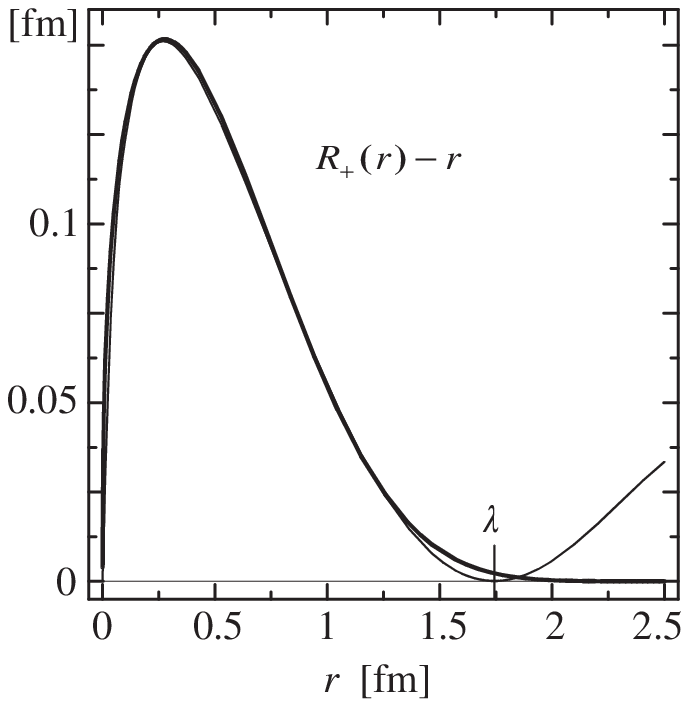,height=65mm}
\end{center}
\mycaption{Exact correlation function 
$\RP{r}$ for the ground state (thin line) and the parameterization
\fmref{E-3-1-A} plotted as $\RP{r}-r$.}{F-3-1-1}
\end{figure} 

\subsection{\element{2}{H} and \element{3,4}{He} with the Malfliet--Tjon V potential}
\label{Sec-3-2}

In this section the particle number dependence of the
correlation function is investigated. For that the three
parameters $\alpha, \beta, \eta$ of the ansatz for $\RP{r}$
given in eq. \fmref{E-3-1-A} are varied in order to minimize the
binding energy of the two--, three-- and four--nucleon
system. The uncorrelated trial states are kept as simple as
possible because the short range correlations are supposed to be
described by the unitary correlator alone. Therefore, the
uncorrelated \element{3}{He} and \element{4}{He} states are
taken to be a product of three and four identical Gaussians in
coordinate space, respectively. The antisymmetry is taken care
of by an appropriate spin--isospin part of the many--body
state. This excludes for \element{3}{He} the admixture of
relative p--states which are important for a correct ground
state energy.

Thus all three uncorrelated many--body wave functions are
characterized by only one variational parameter, which for
\element{2}{H} is $\rho$ as defined in \fmref{E-2-4-E} and for
\element{3,4}{He} is the width $a$ of the Gaussian shaped
single--particle state.

The correlated Hamiltonian is calculated up to its two--body
part as defined in section \xref{Sec-2-1}
\begin{eqnarray}
\coop{H}^{C2} = \op{T}_{int} + \coop{T}^{[2]} + \coop{V}^{[2]}
\end{eqnarray}
where the intrinsic kinetic energy
\begin{eqnarray}
\op{T}_{int} = \op{T} - \op{T}_{cm}
\ ,\quad
\op{T}_{cm}
=
\frac{1}{2 m A} \Big( \sum_{i=1}^A \op{\vek{p}}(i) \Big)^2
\end{eqnarray}
is the kinetic energy minus the centre of mass kinetic energy
$\op{T}_{cm}$. 

\setlength{\tabcolsep}{5pt}
\begin{table}[b]
\begin{center}
\begin{tabular}{c|c c c|c c|c c|c c}
\hline
A & $\alpha$ & $\beta$ & $\eta$ & $a$ & $\rho$ & 
$\erw{\coop{T}_{int}^{C2}}$ & $\erw{\coop{V}^{C2}}$ &
$\erw{\coop{H}^{C2}}$ & exact \\
& [fm] & [fm] & & [fm$^2$] & [fm] &
[MeV] &  [MeV]  & [MeV] & [MeV] \\
\hline\hline
2 & 0.939 & 0.996 & 0.394 & -- & 2.42 & 4.78 & -5.17 &
 -0.40 & -0.41\\
3 & 0.944 & 1.018 & 0.389 & 2.06 & -- &34.97 &-41.49 &
 -6.52 & -8.26$\pm$0.01\\
4 & 0.954 & 1.121 & 0.369 & 1.51 & -- &76.37 &-107.2&
-30.78 & -31.3$\pm$0.2\\
\hline
\end{tabular}
\vspace*{5mm}
\end{center}
\mycaption{Parameters $\alpha, \beta, \eta$ of the correlator and
$a, \rho$ of the uncorrelated many--body state which minimize
the energy $\erw{\coop{H}^{C2}}$.
$\erw{\op{T}_{int}^{C2}} = \erw{ \op{T}_{int} + \coop{T}^{[2]}}$
is the intrinsic correlated kinetic energy and  
$\erw{\coop{V}^{C2}} = \erw{\coop{V}^{[2]}}$ the correlated
potential energy. The exact binding energies are taken from
\cite{ZSK82}.}{T-3-2-1} 
\end{table}
\setlength{\tabcolsep}{6pt}

Table \xref{T-3-2-1} summarizes the result of minimizing the
expectation value of the correlated intrinsic Hamiltonian
$\coop{H}^{C2}$. The first and very important result is that the correlation
functions $\RP{r}$ (represented by their parameters $\alpha,
\beta, \eta$), which result from minimizing the energy, are very
similar for the two--, three-- and four--body system. This can
also be seen in \figref{F-3-2-1} where they are plotted together
with the $\RP{r}$ which results from mapping the uncorrelated
deuteron ground state \fmref{E-2-4-E} to the exact relative wave
function up to $r=\lambda$ as defined in \fmref{E-2-4-F} and
discussed in section \xref{Sec-2-4}.

\begin{figure}[!bt]
\begin{center}
\epsfig{file=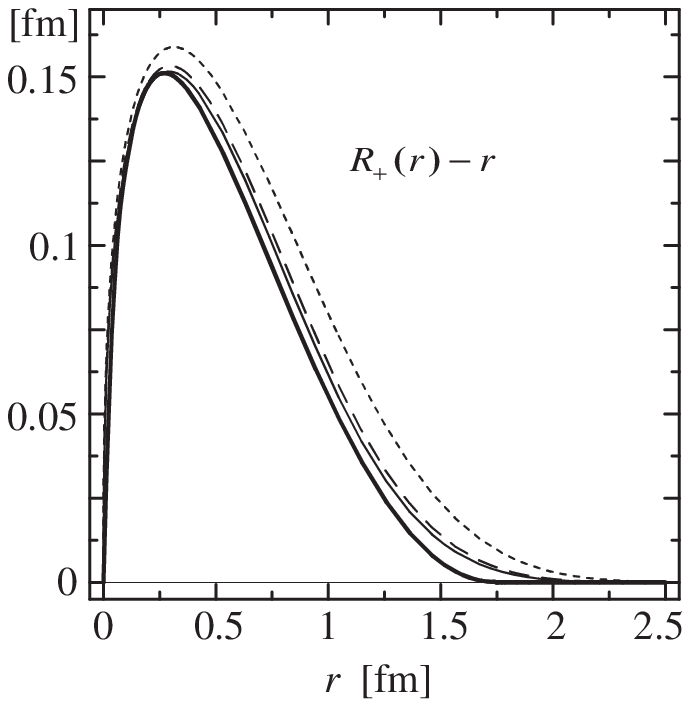,height=65mm}
\end{center}
\mycaption{Comparison between exact correlation function for
the ground state of
\element{2}{H} (thick line) and parameterized functions
obtained by minimizing the ground state energies of 
\element{2}{H} (thin line), \element{3}{He} (long dashed line)
and \element{4}{He} (short dashed line) in terms of
$\RP{r}-r$. }{F-3-2-1}
\end{figure} 

Since all correlation functions turn out to be almost identical, the
initial concept of a unitary state--independent correlator, which
does not depend on the actual system it is applied to, is
strongly supported.

The second result is that despite of the simplistic nature of
the uncorrelated states the binding energies of the $A=2$ and
$A=4$ systems are very close to the exact values. The $A=3$
system is still $1.8\MeV$ above the value of a Green's function
Monte Carlo calculation \cite{ZSK82}. Responsible for that are the
missing exponential tails in the Gaussian radial functions and a
missing component in the uncorrelated state in which the spins of
the two protons are coupled to one and the relative motion is in
a $p$--state.

One should keep in mind that the total energy results from the
subtraction of two large numbers, the kinetic and the potential
energy, both depending sensitively on $\RP{r}$. As will be
demonstrated with an example below, the potential energy almost
doubles when including the correlator.

Since the difference between the correlation functions is so small a
standard correlator with the parameters
\begin{eqnarray}\label{E-3-2-A}
\alpha=0.94\fm\,\quad
\beta=1\fm\ ,\quad
\eta=0.37
\end{eqnarray}
is used for all three systems and the minimization is done only
with respect to a single parameter, $\rho$ or $a$, which
determines the radial wave functions. As expected their values
differ very little from those given in table \xref{T-3-2-1}. The
resulting energies are decomposed into their components in
\figref{F-3-2-2}.

\begin{figure}[!bt]
\begin{center}
\epsfig{file=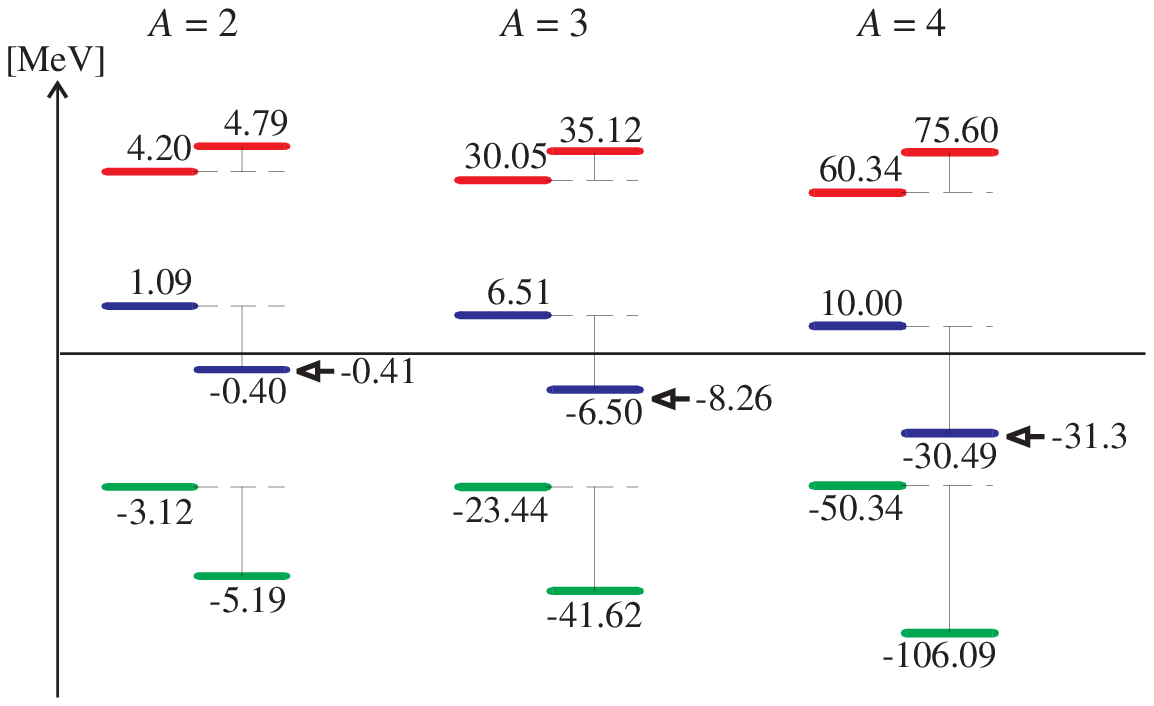,height=65mm}
\end{center}
\mycaption{Energies for $A=2,3,4$ and MTV potential. 
Left columns: expectation values of kinetic energy
$\erw{\op{T}_{int}}$, potential $\erw{\op{V}}$ and 
sum of both $\erw{\op{H}}$. Right columns: expectation 
values of correlated kinetic energy 
$\erw{\op{T}_{int}+\coop{T}^{[2]}}$, correlated potential
$\erw{\coop{V}^{[2]}}$ and sum of both
$\erw{\coop{H}^{C2}}$.
Arrows indicate exact values from \cite{ZSK82}.}{F-3-2-2}
\end{figure} 

For each system the uncorrelated expectation values of kinetic,
potential and total energy are compared with their respective
correlated counter parts. The uncorrelated energies are
calculated with the uncorrelated states which minimize the
correlated energy. In all three cases the kinetic correlation
energy $\erw{\coop{T}^{[2]}}$ (indicated by a vertical line) 
is much less than the potential correlation energy
$\erw{\coop{V}^{[2]}}-\erw{\op{V}}$ so that the sum of both
reduces the uncorrelated total energy from positive values
(unbound) down to almost the exact ground state energies.

\begin{figure}[!bt]
\begin{center}
\vspace*{2ex}
\epsfig{file=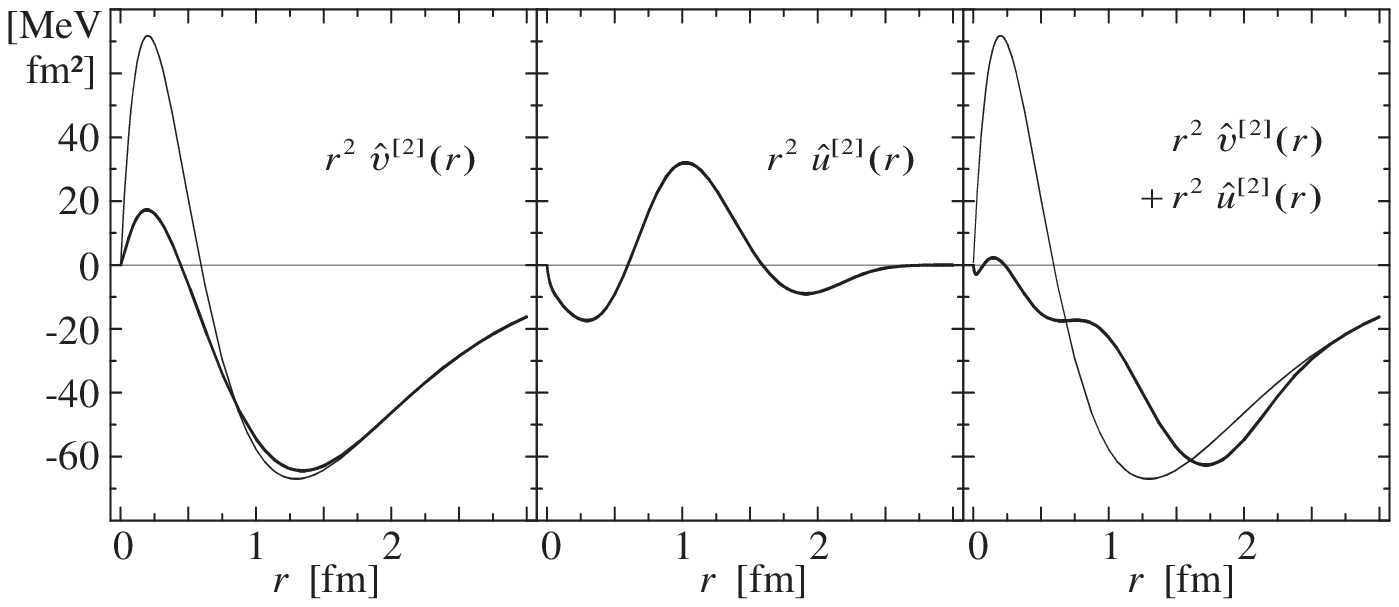,width=135mm}
\end{center}
\mycaption{L.h.s.: correlated
($\coop{v}^{[2]}(r)=\op{v}(\RP{r})$, full line) and uncorrelated
($\op{v}(r)$, thin line) MTV potential.
Centre: potential like part of the correlated kinetic energy
$\coop{u}^{[2]}(r)$.
R.h.s.: MTV potential (thin line) and ``tamed'' potential
$\coop{u}^{[2]}(r)+\coop{v}^{[2]}(r)$.
All potentials are multiplied with $r^2$.}{F-3-2-3}
\end{figure} 

Figure \xref{F-3-2-3} illustrates the action of the unitary
correlator. On the left hand side one sees that the repulsive
maximum of $r^2\op{v}(r)$ is reduced by almost a factor of
three, while the attractive region is enlarged. This corresponds
to the shift of the wave function away from the repulsive 
core towards the attractive region, c.f. \fmref{E-2-2-I}. 
The correlated kinetic energy contributes a potential like part 
$\coop{u}^{[2]}(r)$ which is given in eq. \fmref{E-2-3-A} and depicted
in the middle of \figref{F-3-2-3}. For short distances
$r\leap 1\fm$ it has a pattern opposite to the correlated
potential, attractive at short and repulsive at larger values of
$r$, so that the sum of both
$\coop{u}^{[2]}(r)+\coop{v}^{[2]}(r)$ (r.h.s. of
\figref{F-3-2-3}) almost vanishes in the region of the repulsive
core of the MTV potential. As discussed in section
\xref{Sec-2-4} a uniform uncorrelated state which is mapped by
the unitary correlator $\op{c}$ onto the exact eigenstate of the
Hamiltonian at energy $E$ leads to
$\coop{u}^{[2]}(r)+\coop{v}^{[2]}(r) = E$ for small $r$.
This explains the small values of
$\coop{u}^{[2]}(r)+\coop{v}^{[2]}(r)$ at small distances. The
deviations from a constant $E$ are due to the fact that $\RP{r}$
is parameterized \fmref{E-3-1-A} and the trial state is not
uniform.

The r.h.s. of \figref{F-3-2-3} shows that the repulsive core of
the original potential is completely transformed away by the
unitary correlator. The uncorrelated states feel only a tamed
potential which is purely attractive.

\begin{figure}[!bt]
\begin{center}
\epsfig{file=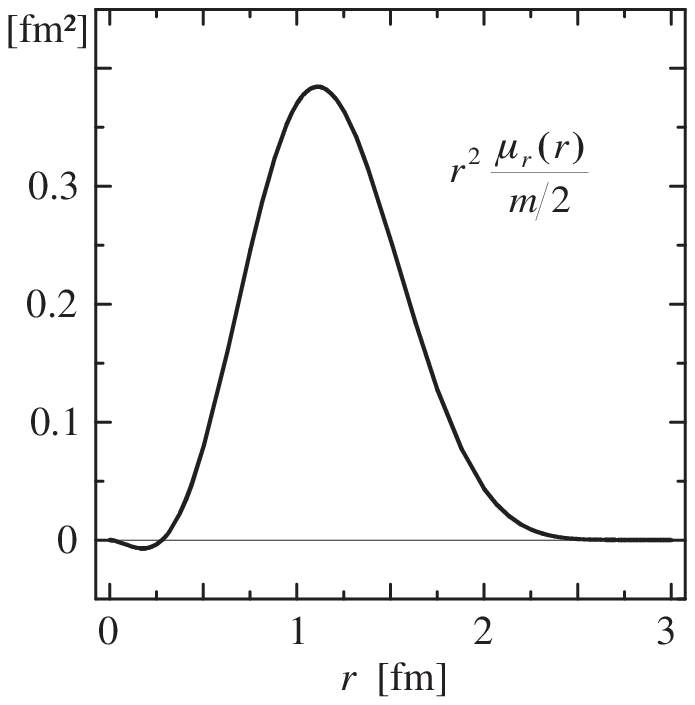,height=65mm}
\hspace{1ex}
\epsfig{file=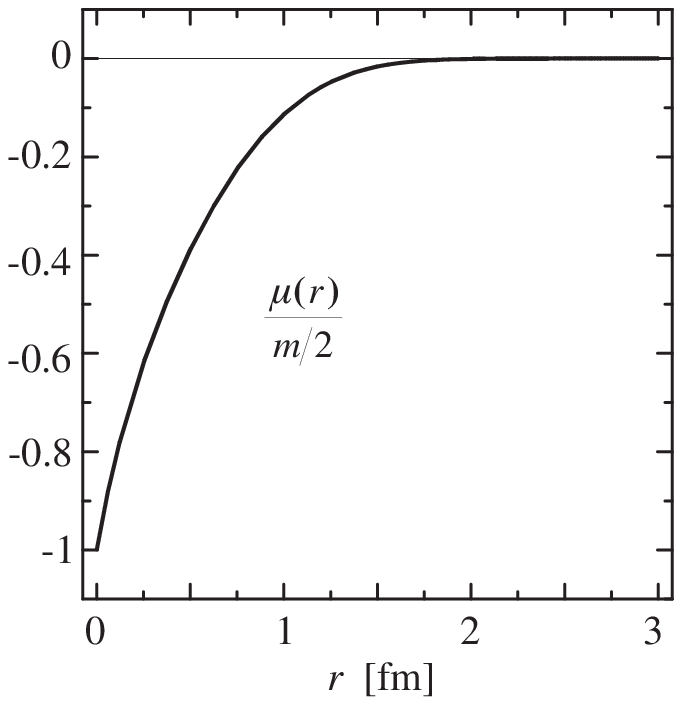,height=65mm}
\end{center}
\mycaption{L.h.s.: Inverse reduced mass in radial direction.
R.h.s.: Inverse relative moment of inertia. The values are
multiplied with $r^2$ and in units of $m/2$.}{F-3-2-4}
\end{figure} 

Since the correlator creates ``holes'' in the many--body state it
induces the additional kinetic energy
$\coop{t}^{[2]}$. One part of that is $\coop{u}^{[2]}(r)$, the
other part can be understood as a momentum dependent two--body
potential which is written in eq. \fmref{E-2-3-AA} in terms of a
reduced mass and a reduced moment of inertia. Their dependence
on the relative distance $r$ is displayed in
\figref{F-3-2-4}. The l.h.s. shows the inverse of the
reduced mass for the radial direction as calculated from
eq. \fmref{E-2-3-D} in units of $m/2$ and multiplied by
$r^2$. It results in an increase of the correlated kinetic
energy. The r.h.s. of \figref{F-3-2-4} depicts the inverse of
the reduced moment of inertia times $r^2$ which is given by
eq. \fmref{E-2-3-E}. For relative angular
momenta $l=0$ this part does not contribute. But for components
with $l\ne0$ it is always negative. This is due to the fact that
the correlator shifts the particles away from each other so that
their moment of inertia increases or the inverse of that
decreases. Therefore, the correlation energy which is the
difference between correlated and uncorrelated energy has to be
negative. Altogether the expectation value of the
momentum--dependent part of
$\coop{T}^{[2]}=\sum_{i<j}\coop{t}^{[2]}(i,j)$ is for $A=4$ only
$3.3$ MeV compared to $12$ MeV for $\erw{\coop{u}^{[2]}}$.

\subsection{Three--body part of the correlated Hamiltonian}
\label{Sec-3-3}

The ansatz for the correlator discussed in section
\xref{Sec-2-0} includes only a two--body generator
$\op{g}^{[2]}(i,j)$. Nevertheless the correlated operators
contain three-- and higher--body operators because 
$\op{g}^{[2]}(i,j)$ appears in the exponent (see
eq. \fmref{E-2-1-D}). 
In principle it is possible that the nuclear many--body system
requires also genuine three--body correlations so that the
corresponding unitary correlator would be written as
\begin{eqnarray}
\op{C}
=
\exp\!\Big(
-i \sum_{i<j}^A \op{g}^{[2]}(i,j)
-i \sum_{i<j<k}^A \op{g}^{[3]}(i,j,k)
\Big)
\ .
\end{eqnarray}
In the $n$--body expansion of a correlated operator
$\coop{B}=\op{C}^{-1}\op{B}\op{C}$ the three--body correlations
induced by $\op{g}^{[3]}$ would appear first in the three--body
part together with the three--body part originating from
$\op{g}^{[2]}$.  It will be very hard or even impossible to
distinguish between the two contributions just from the three--
or four--particle ground state energies.  This uncertainty
together with the complicated calculation of the three--body
contribution of $\op{g}^{[2]}$ as given in
\fmref{E-2-1-D} suggests to estimate the three--body
contribution first before doing elaborate calculations. A small
three--body contribution to $\op{C}^{-1}\op{B}\op{C}$ from
$\op{C}=\exp\{-i\sum_{i<j}^A\op{g}^{[2]}(i,j)\}$ is of course
desirable for justifying the two--body approximation. On the
other hand if it turns out to be small even a small genuine
three--body generator $\op{g}^{[3]}$ might be of equal or more
significance and one would have to discuss its shape and
importance.  Thus the idea is that either it is sufficient to
consider the correlated operators only up to the two--body part
or the method becomes too complicated for practical purposes.

The two--body approximation will be tested by calculating the
correlated Hamiltonian up to three--body operators with the
following approximation which is a partial sum of a power
expansion of $\op{C}^{-1} \op{H} \op{C}$
\cite{Rot97}.
\begin{eqnarray}
&&\op{c}(i,j,k)^{-1} \;
\op{h}(i,j,k) \;
\op{c}(i,j,k)
\\
&&\qquad
\approx
\frac{1}{3!}
\sum_{P\{\alpha,\beta,\gamma\}}
\op{c}(\alpha)^{-1} \,
\op{c}(\beta)^{-1}  \,
\op{c}(\gamma)^{-1} \;
\op{h}(i,j,k) \;
\op{c}(\gamma) \,
\op{c}(\beta) \,
\op{c}(\alpha)
\nonumber
\end{eqnarray}
where $\op{h}(i,j,k) = \op{t}^{[1]}(i) +  \op{t}^{[1]}(j) 
+ \op{t}^{[1]}(k) + \op{v}^{[2]}(i,j) + \op{v}^{[2]}(i,k) + \op{v}^{[2]}(j,k)$.
The greek indices $\alpha, \beta, \gamma$ denote the three
possibilities to pick a pair of particle indices out of $\{i,j,k\}$. 
The summation runs over all $3!$ possibilities to order the 
triple $\{\alpha,\beta,\gamma\}$.
$\op{c}(\alpha)$ is the two--body correlation operator acting on the pair
labeled by $\alpha$.
This approximation correlates all pairs --- so to speak ---
one after the other in all different orderings instead of
simultaneously as $\exp\{-i\sum_{i<j}^A\op{g}^{[2]}(i,j)\}$
would do.

It is not exactly unitary anymore but it conserves the norm in
the sense that the correlated unit operator remains unchanged
\begin{eqnarray}
\coop{1}^{C3} = 1
\ .
\end{eqnarray}
The actual calculations for the three--body and four--body
system are done by Monte Carlo integration. The derivatives of
the kinetic energy are calculated numerically.

For example in the three--body system the potential energy is
given by
\begin{eqnarray}
\erw{\coop{V}^{C3}}
&=&
\frac{1}{3!}
\sum_{P\{\alpha,\beta,\gamma\}}
\int \dint^3x_1 \dint^3x_2 \dint^3x_3\;
\\
&&\quad \times
\sum_{i<j}^3\; \op{v}(\vek{x}_i-\vek{x}_j) \;
|\bra{\vek{x}_1,\vek{x}_2,\vek{x}_3}
\op{c}(\alpha) \,
\op{c}(\beta) \,
\op{c}(\gamma) 
\ket{\Phi}|^2
\nonumber
\end{eqnarray}
and the kinetic energy is calculated as
\begin{eqnarray}
\erw{\coop{T}^{C3}}
&=&
\frac{1}{3!}
\sum_{P\{\alpha,\beta,\gamma\}}
\int \dint^3x_1 \dint^3x_2 \dint^3x_3\;
\\
&&\quad \times
\frac{1}{2m} 
\sum_{i}^3
|\pp{}{\vek{x}_i}
\bra{\vek{x}_1,\vek{x}_2,\vek{x}_3}
\op{c}(\alpha) \,
\op{c}(\beta) \,
\op{c}(\gamma)
\ket{\Phi}|^2
\nonumber
\ .
\end{eqnarray}
The uncorrelated states $\ket{\Phi}$ are those which minimize
the two--body correlated Hamiltonian together with the standard
correlation function $\RP{r}$ defined in section \xref{Sec-3-2}.

\setlength{\tabcolsep}{8pt}
\begin{table}[b]
\begin{center}
\begin{tabular}{c|c c|c c}
\hline
& $A=3$ & Error & $A=4$ & Error\\
\hline\hline
a [fm${}^2$]  &   2.07 &       &   1.55 & \\
\hline
$\erw{\op{T}_{int}}$         & 30.05 &  & 60.39 &\\
$\erw{\coop{T}_{int}^{C2}}$  & 35.12 &  & 75.58 &\\
$\erw{\coop{T}_{int}^{C3}}$  & 35.13 & $\pm0.02$ & 75.56 & $\pm0.06$\\
\hline
$\erw{\coop{T}^{[2]}}$       &  5.07 &  & 15.19 &\\
$\erw{\coop{T}^{[3]}}$       &  0.01 & $\pm0.02$ & -0.02 & $\pm0.06$\\
\hline
$\erw{\op{V}}$               &-23.54 &  &-50.23 &\\
$\erw{\coop{V}^{C2}}$        &-41.62 &  &-106.12&\\
$\erw{\coop{V}^{C3}}$        &-41.48 & $\pm0.001$ &-105.03& $\pm0.002$\\
\hline
$\erw{\coop{V}^{[2]}}$       &-18.08 &  &-55.89 &\\
$\erw{\coop{V}^{[3]}}$       &  0.14 & $\pm0.001$ &  1.09 & $\pm0.002$\\
\hline
$\erw{\op{H}}$               &  6.51 &  & 10.16 &\\
$\erw{\coop{H}^{C2}}$        & -6.50 &  & -30.54&\\
$\erw{\coop{H}^{C3}}$        & -6.35 & $\pm0.02$ & -29.47& $\pm0.06$\\
\hline
$\erw{\coop{H}^{[2]}}$       &-13.01 &  &-40.70 &\\
$\erw{\coop{H}^{[3]}}$       &  0.15 & $\pm0.02$ &  1.07 & $\pm0.06$\\
\hline
\end{tabular}
\vspace*{5mm}
\end{center}
\mycaption{Uncorrelated energies, two-- and three--body 
approximation of correlated energies, all in MeV.}{T-3-3-1} 
\end{table}
\setlength{\tabcolsep}{6pt}

The results are summarized and compared with the two--body
approximation (for which analytical expressions exist) in table
\xref{T-3-3-1}.

The correlated three--body kinetic energy
$\erw{\coop{T}_{int}^{C3}}$ is for both nuclei within the error
bars of the Monte Carlo integration the same as the correlated
two--body kinetic energy $\erw{\coop{T}_{int}^{C2}}$. This means
that the three--body contribution 
$\erw{\coop{T}^{[3]}}=\erw{\coop{T}_{int}^{C3}}-\erw{\coop{T}_{int}^{C2}}$
is within $\pm0.06\MeV$ for $A=4$.

The three--body contribution to the correlated potential 
$\erw{\coop{V}^{[3]}}=\erw{\coop{V}^{C3}}-\erw{\coop{V}^{C2}}$
is of the order of $0.14\MeV$ for $A=3$ and $1.09\MeV$ for
$A=4$. Both corrections are small compared to the difference
$\erw{\coop{V}^{C2}}-\erw{\op{V}}$ between the two--body
correlated and the uncorrelated potential which is $-18.08\MeV$ for 
$A=3$ and $-55.89\MeV$ for $A=4$.

It is interesting to note that the three--body contribution
shifts up the energy again slightly by $2\%$ of the gain
in binding achieved by the two--body part of the correlation. 
However, the uncorrelated state is not varied for the three--body
approximation separately but taken from the energy minimization
within the two--body approximation. A variation could lower the
energy again.

The three--body part of the correlated potential is positive in all cases
we considered. The reason is that a simultaneous action of
$\op{c}(i,j,k)$ on three particles which are close shifts
them further away from each other than the sum of pair wise
shifts $\op{c}(i,j)$, $\op{c}(i,k)$ and $\op{c}(j,k)$. 
As $\op{c}(i,j)$ is optimized to move probability from the
repulsive into the attractive region (see \figref{F-2-4-5}) a
further shift results in a loss of binding.

In order to see the dependence on density of the three--body contribution
to the total kinetic energy and potential energy we calculate 
$\erw{\coop{T}^{[3]}}$ and $\erw{\coop{V}^{[3]}}$ as a function
of the width parameter $a$ for \element{3}{He}.
In \figref{F-3-3-1} the ratios 
$\erw{\coop{T}^{[3]}}/\erw{\coop{T}^{C3}}$ and 
$\erw{\coop{V}^{[3]}}/\erw{\coop{V}^{C3}}$
between the three--body part and the total correlated energies
are plotted versus $\kappa = \rho_{max} V_c$, where $\rho_{max}$
is the maximum of the one--body density and $V_c = 0.16
\text{fm}^3$ is the correlation volume defined in eq.
\fmref{E-2-2-P} for the correlation function (parameters in 
eq. \fmref{E-3-2-A}) used in this section.
While the three--body  part of the correlated kinetic energy
stays rather small the three--body part of the correlated
potential energy grows linearly with $\kappa$.
The ratios stay below $1\%$ for $0 \le \kappa \le 0.1$. 
The ground state values for \element{3}{He} and \element{4}{He}
are in this range. One should however keep in mind that the
error induced in the total energy $\erw{\coop{T}^{C2}} +
\erw{\coop{V}^{C2}}$ by neglecting the three--body contributions
is larger because kinetic and potential energy have opposite
sign and are of similar size.
A rule of thumb is to stay with $\kappa$ below $0.1$ if an
accuracy of less then $1\%$ is desired for the effective potential
in two--body approximation.
 
\begin{figure}[!bt]
\begin{center}
\epsfig{file=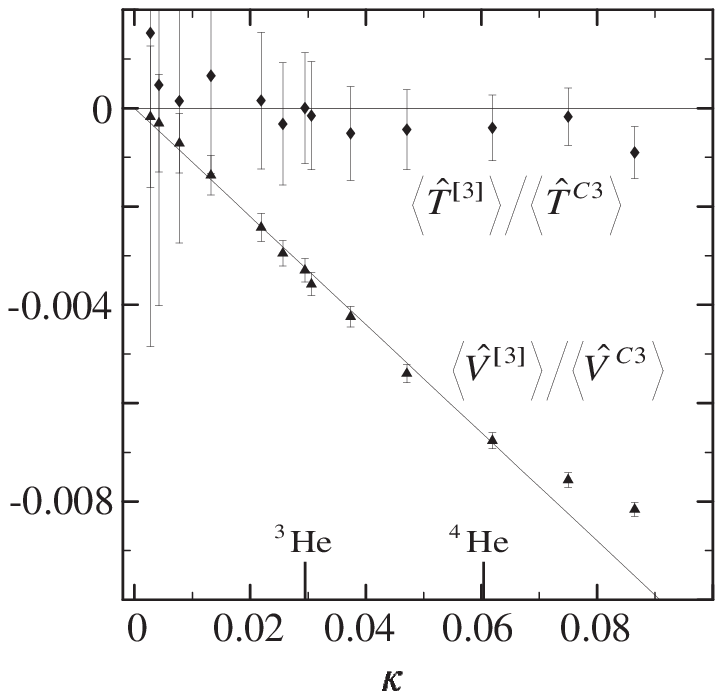,height=65mm}
\end{center}
\mycaption{Ratios of correlated three--body part to total
kinetic $\erw{\coop{T}^{[3]}}/\erw{\coop{T}^{C3}}$ and potential
energy $\erw{\coop{V}^{[3]}}/\erw{\coop{V}^{C3}}$ as function of
$\kappa = \rho_{max} V_c$ ($V_c = 0.16 \text{fm}^3$) for
\element{3}{He}.
Values of $\kappa$ for the ground states of \element{3}{He} and
\element{4}{He} are indicated.
Lines are only guiding the eye. }{F-3-3-1}
\end{figure} 

Since the estimated three--body contribution turns out to be so
small compared to the leading two--body contribution it is not
worthwhile to pursue it further. The other uncertainties, like a
genuine three--body contribution, the parameterization of the
correlation function $\RP{r}$ or the simplicity of the
uncorrelated many--body state are too large.

\section{Many--body systems with spin--isospin--dependent forces}
\label{Sec-4-0}

In this section the unitary correlation operator method is
applied to nuclei up to mass $A=48$. Since the Malfliet--Tjon V
potential overbinds nuclei beyond $A=4$ strongly the so called
modified Afnan--Tang S3 potential \cite{AfT68,GFM81} for which several
calculations exist in the literature \cite{GFM81,KaG92,CFF92,SCF96} is used.
It is specified in the appendix in eq. \fmref{E-A-B}.
The radial dependence of the potential in the two even--parity
states with $S=0$ and $S=1$ and in the odd state is depicted in
\figref{F-4-1-1}.  

The uncorrelated many--body states are Slater determinants
composed of single--particle states
\begin{eqnarray}
\braket{\vek{x}}{q_k}
=
\exp\left\{
-\frac{(\vek{x}-\vek{b}_k)^2}{2 a_k}
\right\}
\otimes \ket{m_{s,k}} \otimes \ket{m_{t,k}} 
\end{eqnarray}
with a Gaussian--shaped coordinate space part, a two--component
spinor \newline
$\ket{m_{s,k}=\pm\half}$ and an isospin part 
$\ket{m_{t,k}=\{\mbox{proton, neutron}\}}$. The complex parameters
$\vek{b}_k$ and $a_k$ are the variational degrees of freedom of
the uncorrelated many--body state $\ket{\Phi}$. They contain the
mean positions, the mean momenta and the complex width of the
packets. For details see ref. \cite{FMDRef}.

All proton and neutron spins are paired to zero except if
the number is odd.

\subsection{Spin--isospin--dependent correlator}
\label{Sec-4-1}

Analogue to the central potential the generator of the unitary
correlator is decomposed into the four spin--isospin channels
\begin{eqnarray}
\op{g}^{[2]}(\vek{r},\vek{q},\vek{\sigma}_1,\vek{\sigma}_2,
\vek{\tau}_1,\vek{\tau}_2)
=
\sum_{S,T}^{\{0,1\}}
\op{g}_{ST}^{[2]}(\vek{r},\vek{q})\;
\Pi_{S}\otimes\Pi_{T}
\end{eqnarray}
with the projection operators 
$\Pi_{S=0} = \frac{1}{4} (1 - \vec{\sigma}_1 \vec{\sigma}_2
)$, $\Pi_{S=1} = \frac{1}{4} (3 + \vec{\sigma}_1 \vec{\sigma}_2
)$ and $\Pi_{T=0}, \Pi_{T=1}$ accordingly. 

The unitary correlator in $A$--body space is herewith
\begin{eqnarray}
\op{C}_A
=
\exp\!\Big\{
-i\sum_{k<l}^A
\op{g}^{[2]}(\vek{r}_{kl},\vek{q}_{kl},\vek{\sigma}_k,\vek{\sigma}_l,
\vek{\tau}_k,\vek{\tau}_l)
\Big\}
\ .
\end{eqnarray}
For the two--body approximation of the correlated operator only
$\op{c} \equiv \op{C_2}$ in the two--body space is needed
\begin{eqnarray}
\op{c}
&=&
\exp\!\Big\{
-i
\sum_{S,T}^{\{0,1\}}
\op{g}_{ST}^{[2]}(\vek{r},\vek{q})\;
\Pi_{S}\otimes\Pi_{T}
\Big\}
\\
&=&
\sum_{S,T}^{\{0,1\}}
\exp\!\big\{
-i \op{g}_{ST}^{[2]}(\vek{r},\vek{q})\;
\big\} \;
\Pi_{S}\otimes\Pi_{T}
\nonumber
\ .
\end{eqnarray}
Since $\Pi_{S}$ and $\Pi_{T}$ are projection operators the
correlator becomes a direct sum of four commuting unitary
correlators, one for each channel.
Compared to the spin--isospin--independent case there are now
four correlation functions $R_{+}^{ST}(r)$ which have to be
determined. 
For the ATS3M potential \cite{AfT68,GFM81} there are only three
because the odd channels $(S,T)=(0,0)$ and $(1,1)$ have the same purely repulsive
radial dependence. The $(S,T)=(0,1)$ and $(1,0)$
potentials have a repulsive core and an attractive tail so that
the procedure discussed in \xref{Sec-2-4} is used to
construct a correlation function of finite range from
eq. \fmref{E-2-4-F} which is then parameterized in the form
\fmref{E-3-1-A}. The resulting parameters are given in table
\xref{T-4-1-1} together with the correlation volumes. The
correlation functions are shown on the r.h.s. of \figref{F-4-1-1}.

\begin{table}[!b]
\begin{center}
\begin{tabular}{m{35mm}|c c c|c}
\hline
\centering channel& $\alpha$ [fm] & $\beta$ [fm] & $\eta$ & $V_c$ [fm${}^3$] \\
\hline\hline
even, $S=0, T=1$ & 1.81 & 1.07 & 0.67 & 0.43 \\
even, $S=1, T=0$ & 1.43 & 0.95 & 0.78 & 0.21 \\
odd, $S=0,\; T=0$\quad\raggedright
odd, $S=1,\; T=1$  & 2.30 & 1.00 & 0.90 & 0.004 \\
\hline
\end{tabular}
\vspace*{5mm}
\end{center}
\mycaption{Parameters of the different correlation functions
$R_{+}^{ST}(r)$ and corresponding correlation volumes $V_c$.}{T-4-1-1} 
\end{table}

\begin{figure}[!bt]
\begin{center}
\epsfig{file=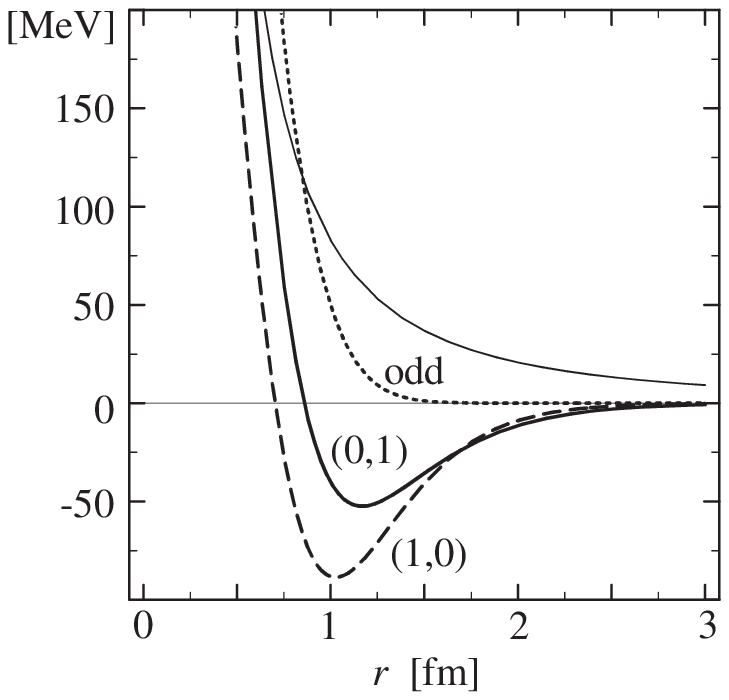,height=65mm}
\hspace{1ex}
\epsfig{file=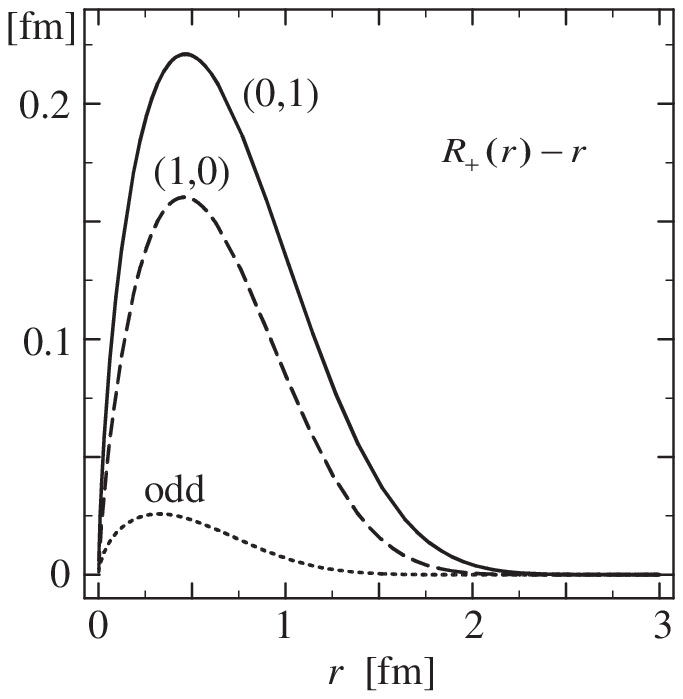,height=65mm}
\end{center}
\mycaption{L.h.s.: Radial dependence of the potentials for even
parity channels $(S,T) = (1,0)$ and $(0,1)$, the odd channels 
$(S,T) = (0,0)$ and $(1,1)$ and centrifugal potential for
$l=1$ (thin line). R.h.s.: Corresponding correlation functions
$R_+^{01}(r)$, $R_+^{10}(r)$,  $R_+^{\text{odd}}(r)$ in terms of
$\RP{r}-r$. }{F-4-1-1}
\end{figure} 

For the repulsive channels $(S,T)=(0,0)$ and $(1,1)$ with an odd
parity in the relative wave function the exact correlation
function obtained by eq. \fmref{E-2-4-F} does not provide a
unique scale parameter $\beta$.  As can be seen from
\figref{F-4-1-1} the centrifugal barrier for $l=1$ states
dominates the repulsive potential for distance $r>0.7$fm and
hence even the uncorrelated $l=1$ state $u(r) = r
j_1(\sqrt{mE}\,r) / \sqrt{N_0} \; \overset{E \to
0}{\longrightarrow} \; r^2 / \sqrt{N_0}$ has already a small
amplitude in the repulsive region.  Therefore, a rather small
correlation function is needed to correct the small distance
behaviour of the repulsive wave function.  $\eta$ which controls
the correlation function at small $r$ is chosen to minimize the
repulsion in the two--body system, while the \element{16}{O}
ground state was employed to choose $\alpha$ and $\beta$.  The
resulting parameters are given in table \xref{T-4-1-1} and the
corresponding $R_+^{\text{odd}}(r) \equiv R_+^{00}(r) =
R_+^{11}(r)$ is depicted on the r.h.s. of \figref{F-4-1-1}.

\subsection{$A=2 \cdots 48$ nuclei with the modified Afnan--Tang force}

The correlation functions given by the parameters of table
\xref{T-4-1-1} are now used to calculate the energies of several
nuclei. The results for \element{4}{He}, \element{12}{C}, 
\element{16}{O}, \element{40}{Ca} and \element{48}{Ca} are
summarized in \figref{F-4-2-1} in terms of the uncorrelated and
correlated kinetic and potential energies as already explained
in connection with \figref{F-3-2-2}. The total energy
$\erw{\coop{H}^{C2}}$ is compared to results of different
methods such as Yakubovsky calculations \cite{KaG92}, FHNC
\cite{CFF92} and CBF \cite{SCF96}.

\begin{figure}[!bt]
\begin{center}
\epsfig{file=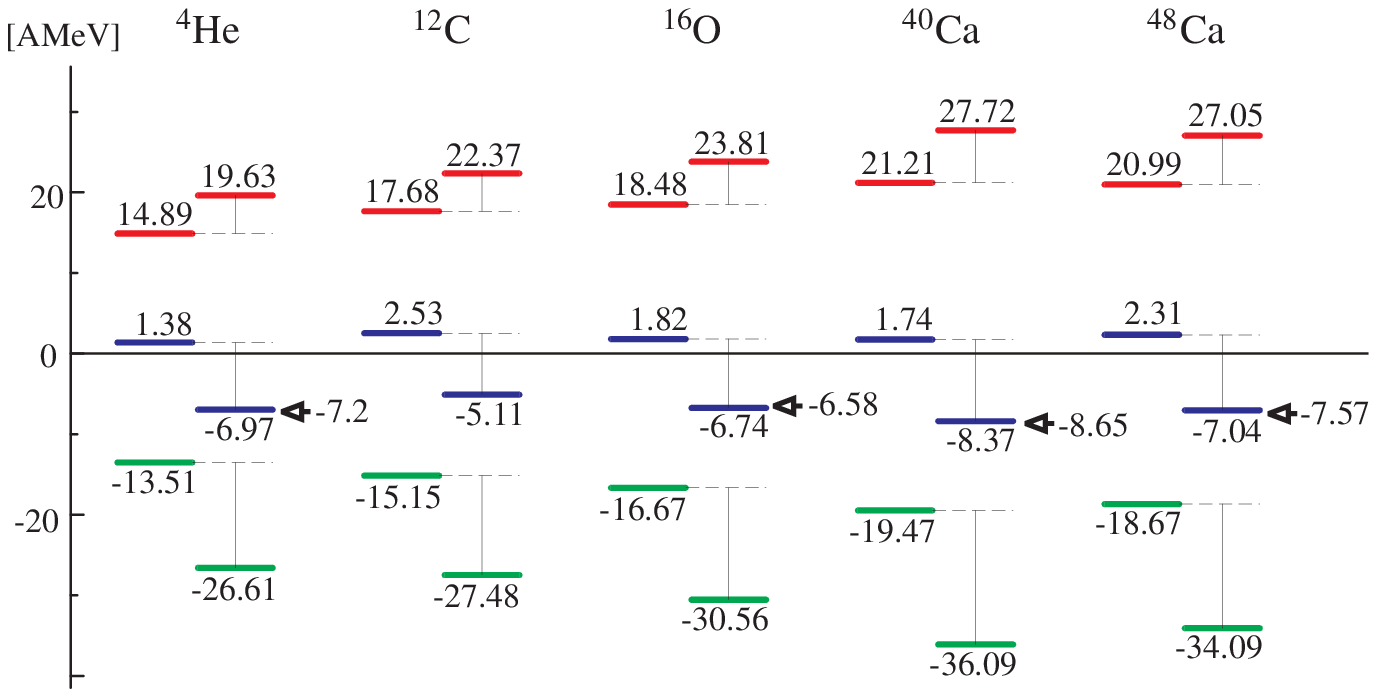,width=135mm}
\end{center}
\mycaption{Uncorrelated and correlated kinetic, potential and total
energies for the modified Afnan-Tang S3 potential, otherwise
same as \figref{F-3-2-2}. Arrows indicate the results of a
Yakubovski calculation for \element{4}{He} \cite{KaG92}, 
of a FHNC calculation for \element{16}{O} \cite{CFF92} and 
of CBF calculations for \element{40,48}{Ca} \cite{SCF96}. }{F-4-2-1}
\end{figure} 

Considering the simplicity of the uncorrelated state and the
fact that the correlated two--body Hamiltonian is the same for
mass number $A=2$ up to $A=48$ the agreement with much more elaborate
methods is striking. For deformed nuclei like
\element{12}{C}, \element{20}{Ne} or \element{28}{Si} the 
variational many--body calculations are very difficult in
terms of computing. With the new Unitary Correlation Operator
Method combined with the uncorrelated states of Fermionic
Molecular Dynamics \cite{FMDRef} deformed nuclei do not pose
special problems. For example for \element{20}{Ne} the 
minimization takes about 3 minutes CPU time on an IBM RS6000, for
\element{40}{Ca} about one hour. 

\begin{table}[!bt]
\vspace*{5ex}
\begin{center}
\begin{tabular}{c|c c c|c|c}
\hline
  & $\erw{\coop{T}_{int}^{C2}}$ & $\erw{\coop{V}^{C2}}$ 
  & $\erw{\coop{H}^{C2}}$ & ``exact'' energy 
  & $\erw{\coop{R_s}^{C2}}^{1/2}$ \\  
\hline\hline
\element{2}{H} & 12.33 & -14.50 & -2.17 
  & -2.22 & 1.93 \\
\element{4}{He} & 78.5 & -106.4 & -27.9 
  & -28.8 \cite{KaG92}; -24.2 \cite{GFM81} & 1.35 \\
\element{6}{Li} & 99.3 & -113.7 & -14.4 
  & & 1.97 \\
\element{12}{C} & 268.5 & -329.8 & -61.3 
  & & 2.36 \\
\element{16}{O} & 381.0 & -488.9 & -107.9 
  & -105.3 \cite{CFF92}; -107.7 \cite{GFM81} & 2.28 \\
\element{19}{F} & 448.2 & -558.8 & -110.6
  & & 2.52 \\
\element{20}{Ne} & 501.1 & -628.4 & -127.4
  & & 2.56 \\
\element{24}{Mg} & 606.7 & -757.6 & -150.9 
  & & 2.74 \\
\element{27}{Al} & 678.2 & -851.1 & -172.9  
  & & 2.82 \\
\element{28}{Si} & 705.4 & -896.1 & -190.7  
  & & 2.88 \\
\element{40}{Ca} & 1108.7 & -1443.6 & -334.9 
  & -346.0 \cite{SCF96}; -335.6 \cite{GFM81} & 2.93 \\
\element{48}{Ca} & 1298.4 & -1636.1 & -337.7 
  & -363.36 \cite{SCF96} & 3.20 \\ 
\hline
\end{tabular}
\vspace*{5mm}
\end{center}
\mycaption{Results of a variational calculation with a single
Slater determinant (except for \element{2}{H}) and a fixed 
correlator for several nuclei with the ATS3M potential. 
Energies are given in MeV, radii in fm.}{T-4-2-1} 
\end{table}

In table \xref{T-4-2-1} ground state energies of a variety of
nuclei, including those of \figref{F-4-2-1}, are listed and
compared to literature values if available.

Considering the fact that for all nuclei the energy results from
the subtraction of a large positive kinetic energy and a
similarly large potential energy, which are about $4$
respectively $5$ times larger than the total energy, it is
surprising that the outcome of the present method lies within
the uncertainties of ``exact'' many--body calculations. 
One reason is probably that the probability in the two--body
density which is  moved out of the repulsive core region is put
where the potential is most attractive, so that a large part of
the attractive correlations are also taken care of.
Otherwise it is hard to imagine that a single Slater determinant
which represents only the mean--field behavior is sufficient as
an uncorrelated state.

The root--mean--square radii of the matter distribution which
are given in the last column of table \xref{T-4-2-1} are
calculated with the correlated operator in two--body
approximation.

\begin{eqnarray}\label{E-4-2-A}
\coop{R_s} 
&&= 
\op{C}^{-1} \frac{1}{A}
\sum_{i}^{A}
\big( \vec{x}(i) - \vec{X}_{cm} \big)^2
\op{C}
= 
\frac{1}{A} 
\big[ 
\op{C}^{-1} \sum_{i}^{A} \vec{x}(i)^2 \op{C}
- \vec{X}_{cm}^2 
\big]
\\
&&\approx \coop{R_s}^{C2} 
= \op{R_s} + \coop{R_s}^{[2]}
\nonumber
=
\frac{1}{A} 
\big[ 
\sum_{i}^{A} \vec{x}(i)^2 -  \vec{X}_{cm}^2 
\big]
+
\frac{1}{2A} 
\sum_{i<j}^{A}
\big( \RP{r_{ij}}^2 - r_{ij}^2 \big)
\nonumber
\end{eqnarray}
where the centre of mass position operator 
\begin{eqnarray}
\vec{X}_{cm} 
= 
\frac{1}{A} \sum_{i}^{A} \vec{x}(i)
\end{eqnarray}
commutes with the correlator $\op{C}$.
The expression \fmref{E-4-2-A} is a straightforward result of
the unitary correlator method explained in section
\xref{Sec-2-0}.

The contribution $\erw{\coop{R_s}^{[2]}}$ of the second order to 
$\erw{\coop{R_s}^{C2}}^{1/2}$ is in all cases only in the last
digit which is displayed.
This shows that the matter distribution in coordinate space is
changed very little by the repulsive two--body correlations. 
\section{Summary}
\label{Sec-5-0}

A new concept, the Unitary Correlation Operator Method (UCOM),
has been developed to describe short range correlations brought
about by the repulsive part of the interaction.

The unitary correlator reduces the probability to find two
particles in the classically forbidden region of the repulsive
core by shifting them away from each other.
The relation between uncorrelated and new correlated
distances are expressed in terms of the coordinate
transformation $r \to \RP{r}$.
With this key quantity, which may depend on spin and isospin,
analytic expressions for the correlated states and the correlated
kinetic and potential energies are derived.
Due to unitarity correlated operators can be evaluated easily,
for example a local potential $v(r)$ between two particles
transforms as $\op{c}^\dagger \, v(r) \, \op{c} = \op{c}^{-1} \,
v(r) \, \op{c} = v( \op{c}^{-1} \, r \, \op{c} ) = v(\RP{r}) $.

For a many--body system the physical challenge is to find 
an optimal $\RP{r}$ which describes well the depletion in the
two--body density at short distances, which cannot be
represented by the uncorrelated states.
Due to unitarity this reduction in probability results in an
increase at larger distances. 
If the two--body potential has an attractive area outside the
repulsion it is natural to enhance the probability there.
These type of repulsive--attractive potentials provide an inherent
length scale for the range of the correlator.

For numerical ease one wants to use the two--body approximation for
correlated operators and states and hence the correlation range
should be smaller than the mean particle distance so that the
probability to find two particles within the correlation volume $V_c$
is small and the chance that three particles are simultaneously 
inside $V_c$ is negligible.
If this condition is not fulfilled because the density is too
high, the correlation range in which $\RP{r}$ differs from $r$
has to be reduced and at the same time the flexibility of the
uncorrelated states has to be increased to include medium range
correlations.
This is an important aspect of the method: the unitary
correlator and the uncorrelated states have to harmonize.

For nuclei (at least without tensor forces which we have not 
investigated yet) this seems to be the case because the
calculations presented show that a single Slater determinant plus
a suitably chosen unitary correlator reproduce the ``exact''
ground--state energies of nuclei between $A = 2$ and $48$ within
a few percent.
\clearpage
{\bf Acknowledgments}\\[5mm]
Part of this work was supported by a grant of the CUSANUSWERK to J.~S..
\appendix
\section{Appendix}
\label{Sec-A}

\subsection{Malfliet--Tjon V potential}

The MTV potential \cite{MaT69} is given by
\begin{eqnarray}\label{E-A-A}
v(r)
=
\sum_{i=1}^2\; \gamma_{i} \frac{\exp\left\{-\beta_{i}\; r\right\}}{r}
\quad\mbox{with}\quad
\begin{tabular}{c|c c}
\hline
$i$ & $\gamma_i$ [MeV fm] & $\beta_i$ [fm$^{-1}$]\\
\hline\hline
1   & -584 & 1.55\\
2   & 1458 & 3.11\\
\hline
\end{tabular}
\ .
\end{eqnarray}

\subsection{Modified Afnan--Tang S3 potential}

The modified Afnan--Tang S3 potential \cite{GFM81} is given by
\begin{eqnarray}\label{E-A-B}
&&\op{v}
=
  v_0(r)\; \op{\Pi}_{01}
+ v_1(r)\; \op{\Pi}_{10}
+ v_{\text{odd}}(r)\; \left(\op{\Pi}_{00}+\op{\Pi}_{11}\right)
\quad\mbox{with}
\\
&&v_S(r)
=
\sum_{i=1}^3\; \gamma_{S,i} \exp\left\{-\beta_{S,i}\;
r^2\right\}
\nonumber
\end{eqnarray}
and the parameters
\begin{table}[hhhh]
\begin{center}
\begin{tabular}{c|c c|c c|c c}
\hline
    & \multicolumn{2}{c|}{$S=0$}
    & \multicolumn{2}{c|}{$S=1$}
    & \multicolumn{2}{c }{odd}\\
$i$ & $\gamma_{S,i}$ [MeV] & $\beta_{S,i}$ [fm$^{-2}$]
    & $\gamma_{S,i}$ [MeV] & $\beta_{S,i}$ [fm$^{-2}$]
    & $\gamma_{S,i}$ [MeV] & $\beta_{S,i}$ [fm$^{-2}$] \\
\hline\hline
1   & 1000 & 3.0 & 1000 & 3.0  & 1000 & 3.0 \\
2   & -166 & 0.8 & -326.7 & 1.05 &        &     \\
3   &  -23 & 0.4 & 1000 & 0.6  &        &     \\
\hline
\end{tabular}
\vspace*{5mm}
\end{center}
\mycaption{Parameters of the modified Afnan--Tang S3 potential.}{T-A-2} 
\end{table}
 
The operators $\op{\Pi}_{ST}=\op{\Pi}_{S}\otimes\op{\Pi}_{T}$
project on two--body states with total spin $S$ and total
isospin $T$, respectively.


\begin{thebibliography}{99}
\bibitem{AHP93}
	A.N.~Antonov, P.E.~Hodgson, I.Zh.~Petkov,
	{\it Nucleon Correlations in Nuclei},
	Springer Series in Nuclear and Particle Physics,
	Springer--Verlag, Berlin (1993)
%
%
\bibitem{Bru55}
	K.A. Brueckner, C.A. Levinson, Phys. Rev. {\bf 97}
	(1955) 1344; 
        K.A. Brueckner, Phys. Rev. {\bf 100} (1955) 36
%
%
\bibitem{BPP93}
	O. Benhar, V.R. Pandharipande, S.C. Pieper,
	Rev. Mod. Phys. {\bf 65} (1993) 817
%
\bibitem{Jas55}
	R. Jastrow,
	Phys. Rev.   {\bf 98}   (1955)  1479
%
%
\bibitem{PrS64}
	J. da Providencia, C.M. Shakin,
	Ann. of Phys. {\bf 30} (1964) 95
%
%
\bibitem{BaK73}
	B.R. Barrett, M.W. Kirson,
	Adv. Nucl. Phys. {\bf 6} (1973) 219
%
%
\bibitem{SOK94}
	K. Suzuki, R. Okamoto, H. Kumagai,
	Phys. Rep. {\bf 242} (1994) 181
%
%
\bibitem{Cla79}
	J.W. Clark,
	Prog. Part. Nucl. Phys. {\bf 2} (1979) 89
%
%
\bibitem{Eks60}
	H. Ekstein,
	Phys. Rev. {\bf 117} (1960) 1590
%
%
\bibitem{CCD70}
	F. Coester, S. Cohen, B. Day, C.M. Vincent,
	Phys. Rev. {\bf C1} (1970) 769
%
%
\bibitem{PaW79}
	V.R. Pandharipande, R.B. Wiringa,
	Rev. Mod. Phys. {\bf 51} (1979) 821
%
%
\bibitem{MoS60}
	S.A. Moszkowski, B.L. Scott,
	Ann. of Phys. {\bf 11} (1960) 65
%
%
\bibitem{MaT69}
	R.A. Malfliet, J.A. Tjon
	Nucl. Phys. {\bf A127} (1969) 161
%
%
\bibitem{VaS95}
	K.~Varga, Y.~Suzuki,
	Phys. Rev. {\bf C52} (1995) 2885
%
%
\bibitem{HeZ85}
	U. Helmbrecht, J.G. Zabolitzky,
	Nucl. Phys. {\bf A442} (1985) 109
%
%
\bibitem{ZSK82}
	J.G. Zabolitzky, K.E. Schmidt, M.H. Kalos,
	Phys. Rev. {\bf C25} (1982) 1111
%
%
\bibitem{Rot97}
	R. Roth, diploma thesis, TH Darmstadt (1997);\\
	(copy from \verb#http://www.gsi.de/~rroth#)
\bibitem{AfT68}
	I.R. Afnan, Y.C. Tang,
	Phys. Rev. {\bf 175} (1968) 1337
%
%
\bibitem{GFM81}
	R. Guardiola, A. Faessler, H. M\"uther, A. Polls,
	Nucl. Phys. {\bf A371} (1981) 79
%
%
\bibitem{KaG92}
	H. Kamada, W. Gl\"ockle,
	Nucl. Phys. {\bf A548} (1992) 205
%
%
\bibitem{CFF92}
	G. Co', A. Fabrocini, S. Fantoni, I.E. Lagaris,
	Nucl. Phys. {\bf A549} (1992) 439
%
%
\bibitem{SCF96}
	F.A. de Saavedra, G. Co', A. Fabrocini, S. Fantoni,
	Nucl. Phys. {\bf A605} (1996) 359
%
%
\bibitem{FMDRef}
	H. Feldmeier,
	Nucl. Phys.   {\bf A515}   (1990)  147;\
	H. Feldmeier, K. Bieler, J. Schnack,
	Nucl. Phys.   {\bf A586}   (1995) 493;\
	H. Feldmeier, J. Schnack,
	Nucl. Phys.   {\bf A583}   (1995) 347;\
	J.Schnack, H. Feldmeier,
	Nucl. Phys. {\bf A601} (1996) 181
%
%
\end{thebibliography}
\end{document}